\documentclass[letter]{aa}
\usepackage[varg]{txfonts}
\usepackage{graphicx}
\usepackage{natbib}
\usepackage{appendix}
\usepackage{amsmath}
\usepackage{mathtools}
\usepackage{enumitem} 
\usepackage{hyperref}
\usepackage{pdflscape}
\usepackage{caption}
\usepackage{rotating}
\usepackage{float}
\usepackage{subfig}
\usepackage{stfloats}

\usepackage{listings}
\usepackage{xcolor}

\hypersetup{
     colorlinks   = true,
     allcolors    = blue
}
\begin{document}

\titlerunning{Benchmarking pre-main sequence stellar evolutionary tracks using disk-based dynamical stellar masses}
\title{Benchmarking pre-main sequence stellar evolutionary tracks using disk-based dynamical stellar masses
  } 

\author{Luigi Zallio\inst{1,2}
  \and Miguel Vioque\inst{2}
  \and Sean M. Andrews\inst{3}
  \and Aaron Empey\inst{4}
  \and Giovanni P. Rosotti\inst{1}
  \and Anna Miotello\inst{2}
  \and Carlo F. Manara\inst{2}
  \and John M. Carpenter\inst{5}
  \and Dingshan Deng\inst{6}
  \and Nicolás T. Kurtovic\inst{7,8}
  \and Charles J. Law\inst{9}\fnmsep\thanks{NASA Hubble Fellowship Program Sagan Fellow}
  \and Cristiano Longarini\inst{10}
  \and Teresa Paneque-Carreño\inst{11}
  \and Richard Teague\inst{12}
  \and Marion Villenave\inst{13}
  \and Hsi-Wei Yen\inst{14}
  \and Francesco Zagaria\inst{8}
  }

\institute{Affiliations can be found at the end of the paper (page 5).}

\date{Received on November 14, 2025 / Accepted on March 3, 2026}

\abstract{Stellar masses are a fundamental property to understand models of pre-main sequence evolution, but their values derived from Hertzsprung–Russell (HR) diagrams are strongly model dependent. We benchmark pre-main sequence stellar evolutionary tracks using stellar masses dynamically estimated by fitting a parametric model to ALMA observations of the $^{12}$CO $(J=3-2)$ line transition emitted by the disks orbiting 20 sources in the old ($4-14$ Myr) Upper Scorpius star forming region. We derive stellar masses from HR diagram fitting for ten different stellar evolutionary models, which we then compare with their stellar dynamical masses for comparison in the stellar mass range $0.1-1.3 \> M_\odot$. Models with a moderate-to-low fraction of cold stellar spots ($f=17\%$) most accurately reproduce the dynamical stellar masses ($100\%$ of the targets agree within $\pm1\sigma$). While a higher spot coverage ($f=34\%$) provides similar stellar mass predictions similar to magnetic equipartition models, larger fractions ($f\geq51\%$) significantly disagree with dynamical masses. Magnetic equipartition models overestimate stellar masses up to a factor $\sim20\%$, whereas non-magnetic models underestimate them up to $\sim12\%$. For some models, there is evidence that the stellar mass discrepancies are anticorrelated with dynamical stellar masses. When stellar dynamical mass priors are considered in HR diagram fitting, the median age of a single source can change up to $\sim25\%$, while the median ages inferred across different tracks become consistent, with the age scatter decreasing by $\gtrsim77\%$. These results provide strong empirical constraints for testing and developing evolutionary models of pre-main sequence stars.}
\maketitle

\section{Introduction}
\label{sec:introduction}

Theoretical models of star formation and pre-main sequence (PMS) evolution are fundamental for deriving the physical properties of Young Stellar Objects (YSOs). The positions of YSOs on the Hertzsprung-Russell (HR) diagram provide the main pathway to estimate stellar masses and ages for entire populations. These quantities are central to studies of star and planet formation and the evolution of protoplanetary disks (e.g., \citealp{2000ApJ...540..255P,2008ASPC..384..200H,2011ApJ...738..140H,2012ApJ...746..154P,2022ApJ...930...39V,Manara_2023,Ratzenbock_2023a}).
Accurate YSO masses are also important on large samples, since several fundamental parameters, such as mass accretion rate and protoplanetary disk mass, scale with stellar mass (\citealp{2003ApJ...592..266M,Andrews_2013,Ansdell_2017}) and are required to derive accretion rates themselves (\citealp{2016ARA&A..54..135H}).
Ages are particularly difficult to constrain (\citealp{2014prpl.conf..219S}) and, when inferred from evolutionary models, are strongly model dependent. More accurate ages from theoretical models would enable direct comparisons with other timescale indicators, thereby improving our understanding of the dynamics and physics of the processes involved in star formation (e.g., Lithium depletion or cluster expansion, \citealp{2022A&A...659A..85F} and \citealp{2024NatAs...8..216M}, respectively). 

Since the seminal works of \citet{1961PASJ...13..450H} and \citet{1965ApJ...142..841H}, PMS models have progressively incorporated a wide range of physical components, including convection, atmospheric evolution, accretion history, rotation, magnetic fields, opacities, and cold stellar spots. Benchmarking these models with independent measurements is needed to improve their physics and define the stellar mass and age ranges where they work best (e.g., \citealp{Simon_2000,Simon_2017,Simon_2019,Rosenfeld_2012,Guilloteau_2014,Czekala_2016,Yen_2018,Sheehan_2019}). In this work we compare stellar masses derived from protoplanetary disk rotation, i.e. ``dynamical'' masses independent of stellar evolutionary models, with stellar masses predicted by modern PMS evolutionary models using observed effective temperatures and luminosities. Specifically, we test the models of \cite{Baraffe2015}, \cite{Feiden2016} (non-magnetic and magnetic), PARSEC v2.0 (\citealt{Nguyen_2022}), \cite{Siess_2000}, and SPOTS (\citealt{Somers_2020}, with different cold photospheric spot coverages). This comparison provides a direct benchmark of their predictive power and the relevance of their underlying physics.

\vspace{-0.6cm}
\section{Sample}
\label{sec:data}

The targets were selected based on the availability of accurate dynamical masses, reliable stellar parameters, and precise \textit{Gaia} parallaxes ($\varpi/\sigma(\varpi)>10$). The parent sample comprises 37 disks in the Upper Scorpius region from \citet{Carpenter2025}, for which \cite{Zallio_2026} measured dynamical stellar masses modeling ALMA Band 7 $^{12}$CO $J = 3-2$ visibilities. 

We excluded from the sample four disks with poorly constrained rotational profiles, and two multiple systems. We then excluded from our analysis four sources for which the moderate angular resolution prevented reliable constraints on the height of the CO emitting layer (the red sources in Fig. \ref{fig:M*_comparison}), together with a cloud-absorbed disk with an inconsistent dynamical mass ($M_{\star,\mathrm{dyn}}\sim1.4\> M_\odot$, but spectral type M2). The final sample is composed of 20 sources from the sample above that have well-determined stellar properties, which were derived from broad-band flux-calibrated medium-resolution spectra (Empey et al., in prep.) derived from VLT/X-Shooter and analyzed with \verb|FRAPPE| (\citealp{Claes_2024}, based on \citealp{Manara_2013}). These spectra are well-suited for young stars because they allow a better modeling of veiling, extinction, and spectral type for the targets (e.g., \citealp{Herczeg_2014, Manara_2013, Manara_2020, Manara_2023}).

\vspace{-0.4cm}
\section{Derivation of dynamical and evolutionary model–based stellar masses}\label{sec:analysis}

\vspace{-0.1cm}
The $^{12}$CO $J=3-2$ visibility modeling is described in detail in \cite{Zallio_2026}. In brief, the analysis involved fitting a parametric model to ALMA visibilities using the \verb|csalt| software\footnote{\url{https://github.com/seanandrews/csalt}} (Andrews et al., in prep.), and sampling the posterior distribution with \verb|emcee| (\citealt{Foreman_2013}). A key free parameter in these models is the stellar mass, which sets the rotational velocity field of the disk and is thus directly constrained by the data. Since the statistical uncertainties reported from \cite{Zallio_2026} are unrealistically small, in Appendix \ref{appendix:masses} we describe how we assign uncertainties to these dynamical stellar mass measurements, denoted $M_{\star,\text{dyn}}$.
\begin{figure*}[t!]
    \centering
    \includegraphics[width=0.8\linewidth]{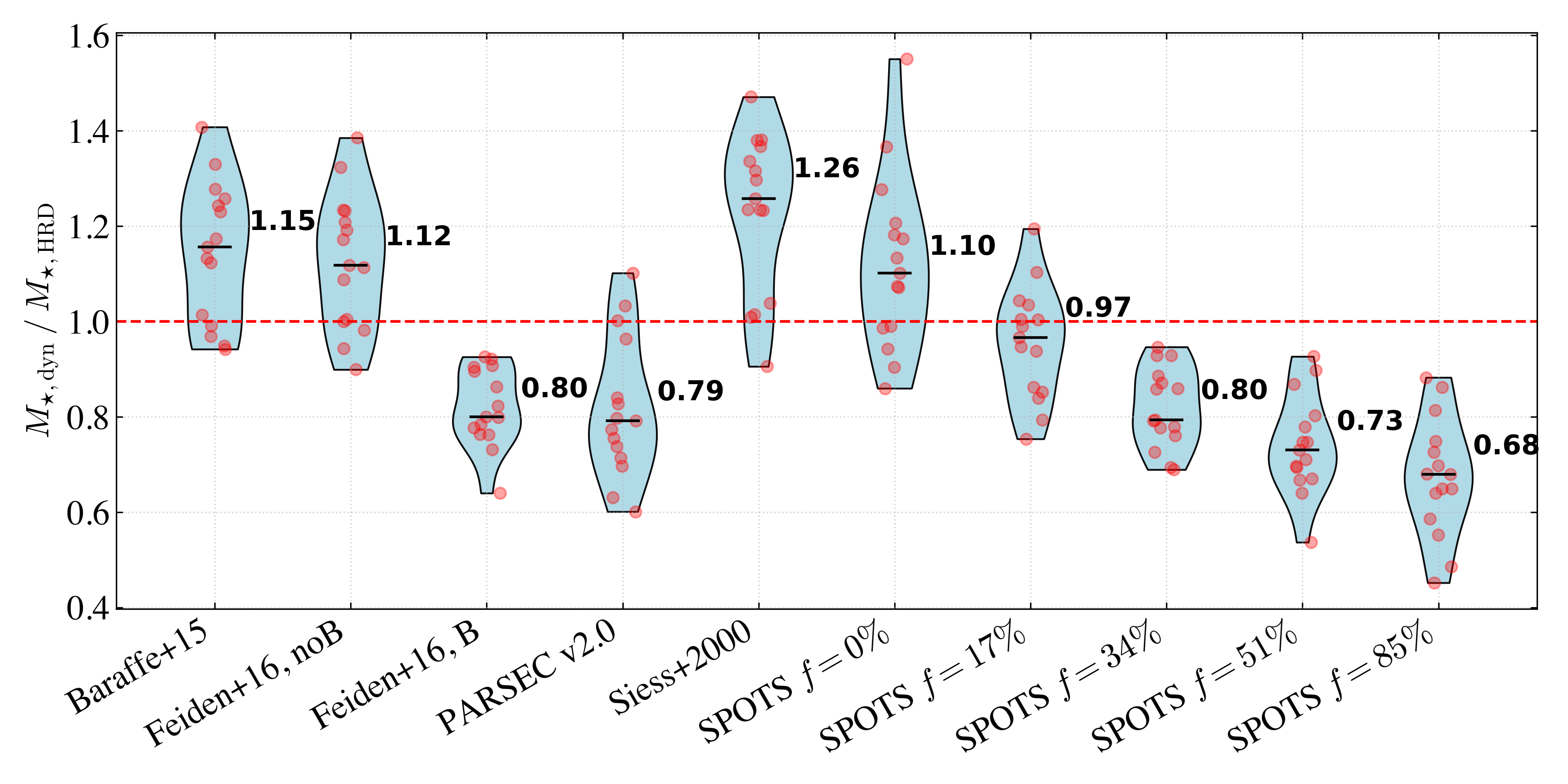}
    \vspace{-0.4cm}
    \caption{Violin plot of the ratio of the dynamical stellar masses from disk rotation ($M_{\star,\text{dyn}}$) and the stellar masses from evolutionary models ($M_{\star,\text{HRD}}$) in the mass range $0.1-1.3 \>M_{\odot}$. One-to-one line is indicated with a red dashed line. The median value of $M_{\star,\text{dyn}} / M_{\star,\text{HRD}}$ is reported next to each distribution.}
    \label{fig:violin}
    \vspace{-0.5cm}
\end{figure*}
In addition to these dynamical masses, we derive stellar masses from the HR diagram ($M_{\star,\text{HRD}}$) using the effective temperatures $(T_{\text{eff}})$ and stellar luminosities $(L_\star)$ presented in Empey et al. (in prep.) for different PMS evolutionary models. In particular, we used the evolutionary models presented in \citealt{Baraffe2015}, the magnetic\footnote{As reported by the author on the \texttt{GitHub} page, the magnetic tracks available on \url{https://github.com/gfeiden/MagneticUpperSco} are incorrect for ages $\leq1$ Myr, containing numerical artifacts, which we corrected by interpolating the data and removing the outliers.} and non-magnetic tracks presented in \cite{Feiden2016}, the PARSEC v2.0 tracks presented in \cite{Nguyen_2022}, the tracks from \cite{Siess_2000}, and the stellar SPOTS tracks presented in \cite{Somers_2020} with five different cold stellar spots fractional coverages of the stellar surface ($f=0\%$, $f=17\%$, $f=34\%$ $f=51\%$, and $f=85\%$). To extract stellar masses and ages (together with their uncertainties), we used the \verb|python| package \verb|ysoisochrone| developed by \cite{Deng_2025}, based on the \verb|IDL| code developed by \cite{Pascucci_2016}, which uses a Bayesian inference approach.
The HR grids used in this work are publicly available in the last version of \verb|ysoisochrone|\footnote{\url{https://github.com/DingshanDeng/ysoisochrone}; the specific version (v1.3.3) used in this work is archived on Zenodo at \url{https://zenodo.org/records/18420752}}. More details and the HR diagrams for each model are presented in Appendix \ref{appendix:hrds} (see Fig. \ref{fig:hrd_spots_085}). In Table \ref{Table:masses_total} we show the collection of stellar masses used and derived in this work.

\vspace{-0.4cm}
\section{Comparison of $M_{\star,\text{dyn}}$ and $M_{\star,\text{HRD}}$}
\label{sec:results}

\begin{figure*}[t]
    \centering
    \includegraphics[width=0.9\linewidth]{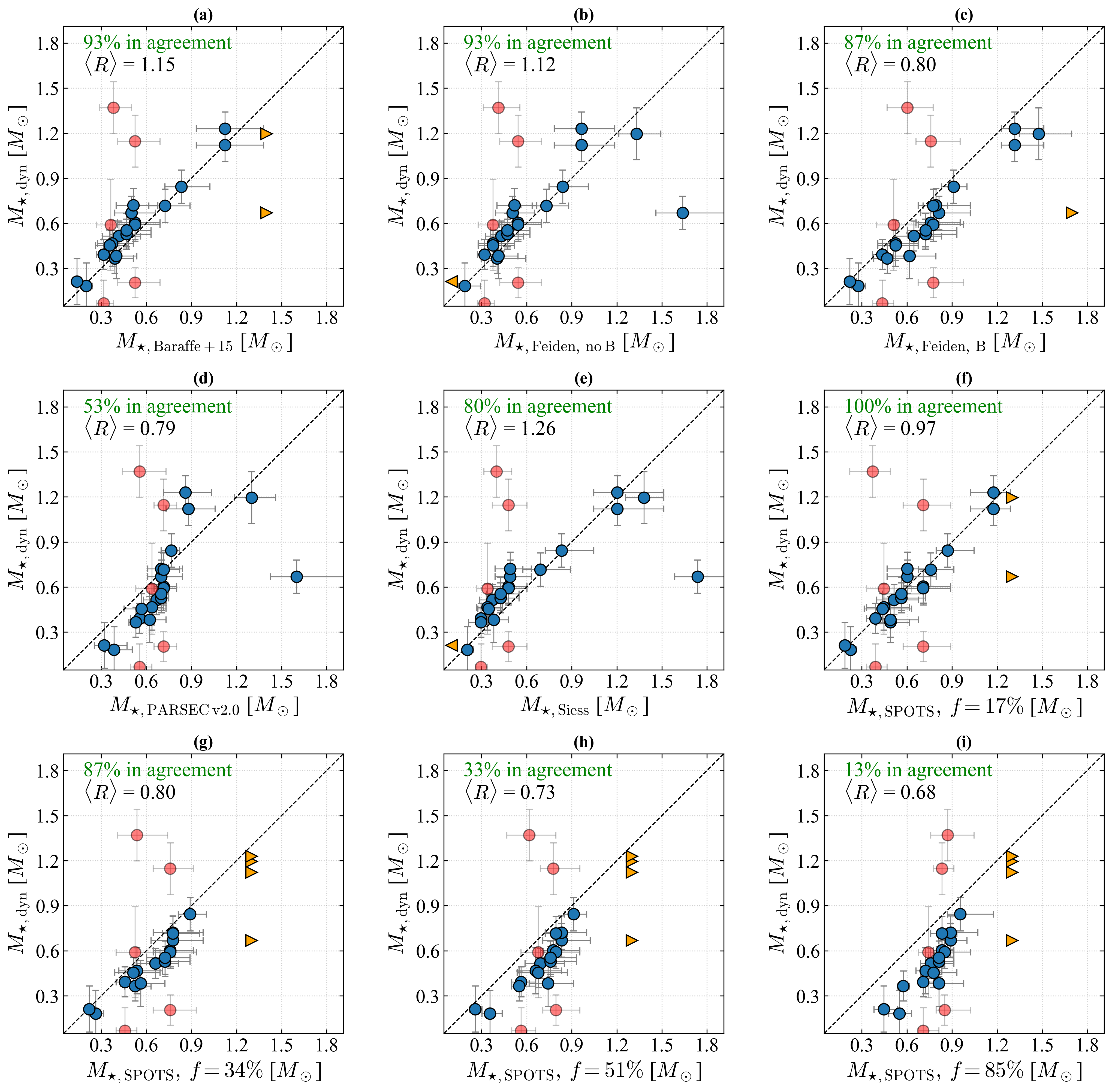}
    \vspace{-0.3cm}
    \caption{Comparison between the dynamical stellar masses $M_{\star,\text{dyn}}$ and the masses from HR diagram fitting $M_{\star,\text{HRD}}$. The red points represent the excluded sources discussed in Sect. \ref{sec:data}, while the orange triangles represent mass upper and lower limits.
    }
    \label{fig:M*_comparison}
    \vspace{-0.5cm}
\end{figure*}

\vspace{-0.1cm}
In Figs. \ref{fig:violin} and \ref{fig:M*_comparison} we compare the dynamical stellar masses ($M_{\star,\text{dyn}}$) with the different evolutionary model-based stellar masses ($M_{\star,\text{HRD}}$). 
The fraction of sources with consistent mass estimates is reported as a percentage (see Fig. \ref{fig:M*_comparison}), providing a quantitative measure of consistency between the dynamical stellar masses and the stellar masses provided by each evolutionary model.
In addition to the fraction of sources in agreement, we quantify the typical offset between dynamically-derived stellar masses and theoretical track-inferred stellar masses by computing the median mass ratio: $\langle R \rangle = \langle M_{\star,\text{dyn}} / M_{\star,\text{HRD}} \rangle$ where the median is taken over all sources included in the comparison. This ratio provides a robust, dimensionless measure of systematic differences: $\langle R \rangle \sim 1$ indicates that the evolutionary model predictions are consistent with the dynamical masses, while values smaller than or greater than unity indicate systematic over- or underestimation, respectively (see Figs. \ref{fig:violin}, \ref{fig:M*_comparison}).

Each pre-main sequence evolutionary model covers a different mass range, as illustrated in the HR diagrams of Fig. \ref{fig:hrd_spots_085}. To ensure a fair comparison among the models, we report the percentage of agreement and the median mass ratio $\langle R \rangle$ only for sources with a defined mass value in all evolutionary models (i.e., within the mass range $0.1 < M_\star < 1.3 \> M_\odot$). We also evaluated the fraction of sources with consistent mass estimates and $\langle R \rangle$ including upper and lower limits, and found that, apart from minor variations, the results remain unchanged.

\vspace{-0.4cm}
\section{Results}

\vspace{-0.1cm}
Figs. \ref{fig:violin} and \ref{fig:M*_comparison} show that the best match with dynamical stellar masses is provided by the SPOTS pre-main sequence tracks with $f=17\%$ cold spot coverage ($100\%$ agreement within $\pm1\sigma$, where $\sigma$ is the quadratic sum of the model and dynamical mass uncertainties) in the stellar range $0.1-1.3$ $M_{\odot}$. Indeed, several authors (e.g., \citealt{Grankin_2008}, \citealt{Gully-Santiago_2017}, \citealt{Gangi_2022}, \citealt{Facundo_2023, Facundo_2024}) invoked moderate to extreme filling factors ($f\sim17 - 90\%$) to model the spectra of Class II sources. \citet{Fang_2025} concluded that $f\sim34 - 51\%$ spot coverage was needed to achieve consistent age estimates across spectral types. We find that $f=17\%$ works better at predicting stellar masses than higher fractional spot coverages. SPOTS models with $f=17\%$ (constant for all the mass range of considered YSOs) predict stellar masses that are in between the measurements of \cite{Feiden2016} non-magnetic and magnetic tracks. When the stellar spot fraction increases (e.g., $f=51\%$, $f=85\%$; Fig. \ref{fig:M*_comparison}), the percentage of agreement decreases ($33\%$ for $f=51\%$, and $13\%$ for $f=85\%$). The SPOTS models with $f=0\%$ behave similarly to the tracks of \cite{Baraffe2015} and \cite{Feiden2016} non-magnetic (see Fig. \ref{fig:violin}), and have a percentage of agreement of $87\%$, similar to that of the \cite{Feiden2016} magnetic tracks.

The tracks of \cite{Baraffe2015} and \cite{Feiden2016} non-magnetic tracks give almost-identical mass measurements, and the same percentage of agreement ($93\%$), though both underestimate the dynamical stellar masses by $15\%$ and $12\%$, respectively. When considering \cite{Feiden2016} magnetic tracks, the percentage of agreement slightly decreases to $87\%$, and the evolutionary model-based stellar masses tend to overestimate the masses derived from disk rotation up to $20\%$.

Other works in the literature have compared the \cite{Feiden2016} magnetic and non-magnetic tracks. For example, \cite{Simon_2019} reported that the magnetic tracks of \cite{Feiden2016} work better than the non-magnetic ones in the stellar range $0.4-1.0$ $M_{\odot}$. \cite{Braun_2021} reported instead that the magnetic tracks work better than the non-magnetic ones in the stellar range $0.6-1.3$ $M_{\odot}$. Moreover, \cite{Towner_2025} recently showed that the magnetic tracks of \cite{Feiden2016} work best for stellar masses $M_\star\leq 1$ $M_{\odot}$, while the non-magnetic ones underestimate the dynamical masses by $\gtrsim25\%$. Our findings suggest that, on average, the agreement in the stellar range $0.1-1.3$ $M_{\odot}$ is similar among the two, but on average the magnetic tracks overestimate the dynamical masses by $\sim20\%$, while the non-magnetic ones underestimate the formers by $\sim 12\%$.

The evolutionary model-based stellar masses from the PARSEC v2.0 tracks show an agreement of $53\%$ with the masses from disk rotation, lower than what is found for other tracks. This result is expected, as the PARSEC v2.0 tracks were first developed for evolved and massive stars, and were then extended towards lower stellar mass values. The tracks of \cite{Siess_2000} give an agreement of $80\%$, worse than \cite{Baraffe2015} and \cite{Feiden2016}, but better than those of PARSEC v2.0, and underestimate the dynamical stellar masses by $\sim26\%$.

To assess whether the mass discrepancies exhibit a stellar-mass dependence, we computed Spearman rank correlation coefficients between $(M_{\star,\rm{HRD}} - M_{\star,\rm{dyn}})/M_{\star,\rm{dyn}}$ and $M_{\star,\rm{dyn}}$. Most models show no statistically significant correlation ($p>0.05$), but PARSEC v2.0 and SPOTS models with $f\geq51\%$ display a negative correlation (Spearman coefficient $\rho\leq-0.77$, with $p<10^{-3}$), implying a significant mass-dependent bias.

One limitation of our analysis lies in the assumption of purely Keplerian rotation (including the dependence on the disk's vertical height) within protoplanetary disks. This simplification neglects the impact of pressure gradients and the disk's self-gravity, which can modify the velocity field. For instance, \cite{Andrews_2024} and \cite{Longarini_2025} demonstrated that deviations from Keplerian motion induced by pressure support can alter the inferred mass of the central object by as much as $5-10\%$. However, due to the limited angular ($0.1-0.3^{\prime\prime}$) and spectral ($\sim480$ ms$^{-1}$) resolution of the data presented in \citet{Carpenter2025}, we emphasize that such deviations remain undetectable, and were therefore not considered in the analysis.

In Appendix \ref{appendix:mass_priors} we demonstrate how age estimates can be improved by prior knowledge of dynamical stellar masses. The comparison shown in Fig. \ref{fig:mass_prior} illustrates that incorporating mass priors from the analysis of \cite{Zallio_2026} can shift the ages inferred from HR diagram fitting with the SPOTS $f=17\%$ tracks by approximately $25\%$ when considering single sources, a result consistent with previous findings (e.g., \citealt{Rosenfeld_2012}). In Fig. \ref{fig:ages} we show the distribution of inferred ages for the Upper Scorpius sources from each set of evolutionary tracks considered in this work, reporting the median value and highlighting how the choice of model leads to significantly different median age estimates. Finally, in Fig. \ref{fig:ages_prior}, we show that when a prior on the stellar mass is considered for HR diagram fitting, the median inferred ages across different tracks become much more consistent, and the scatter among different evolutionary tracks reduces from $3.4$ Myr to $0.8$ Myr, an improvement $\gtrsim77\%$.

\vspace{-0.6cm}
\section{Conclusions}
\label{sec:conclusions}

\vspace{-0.1cm}
By comparing accurate measurements of $^{12}$CO disk rotation with stellar masses derived from HR diagrams, we conclude:
\begin{enumerate}[label=(\roman*),itemsep=1pt, topsep=1pt, parsep=1pt, partopsep=1pt]
    \item The SPOTS evolutionary tracks of \cite{Somers_2020} with $f=17\%$ best reproduce ($100\%$ agreement within $\pm1\sigma$ uncertainties) the dynamical stellar masses in the stellar range $0.1-1.3$ $M_{\odot}$.
    \item On average, the HR diagram masses derived using the magnetic tracks of \cite{Feiden2016}, PARSEC v2.0 (\citealt{Nguyen_2022}), and SPOTS (\citealt{Somers_2020}) with $f=34\%$, $51\%$, $85 \%$ overestimate the dynamical masses by $\sim20\%$, $\sim21\%$, $\sim20\%$, $\sim27\%$, $\sim32\%$, respectively. Those provided by the tracks of \cite{Baraffe2015}, \cite{Feiden2016} non-magnetic, \cite{Siess_2000} and SPOTS with $f=0\%$ underestimate them by $\sim15\%$, $\sim12\%$, $\sim26\%$, $\sim10\%$, respectively.
    \item When using stellar dynamical masses as mass priors for HR diagram fitting, the median inferred ages across different tracks become consistent, and the scatter decreases by $\gtrsim77\%$, while individual ages can vary by up to $\sim25\%$. For our sample, we evaluate a median age of $6.8$ Myr, with a scatter of $0.8$ Myr across the ten evolutionary tracks considered.
\end{enumerate}
Since evolutionary models have a strong time-dependent aspect, our comparison demonstrates the consistency of disk-based stellar masses with HR diagram predictions at the age of Upper Scorpius ($4-14$ Myr, \citealp{Ratzenbock_2023a,Ratzenbock_2023b}). Whether this agreement holds at different ages remains to be tested, and will be explored, for example, with the DEC/O Large Program for the younger ages.

\vspace{-0.4cm}
\section*{Data availability}
\label{sec:data_avail}
Table \ref{Table:masses_total} is available in electronic form at the CDS via anonymous ftp to cdsarc.u-strasbg.fr (130.79.128.5) or via \url{http://cdsweb.u-strasbg.fr/cgi-bin/qcat?J/A+A/}, and on \texttt{GitHub}\footnote{\label{footnote:github}\url{https://github.com/lzallio/Benchmarking_PMS_tracks}}.

\vspace{-0.1cm}
\begin{acknowledgements}
We thank the anonymous referee for their careful reading of the manuscript and constructive suggestions. L.Z., M.V. and G.R. are grateful to F. Pérez-Paolino and J. Bary for their important suggestions on pms evolutionary tracks, and to C. Pincon and A. Bressan for their feedback on interpreting the age distributions predicted by different evolutionary models. L.Z. and G.R. acknowledge support from the European Union (ERC Starting Grant DiscEvol, project number 101039651), and from Fondazione Cariplo, grant No. 2022-1217. C.F.M. is funded by the European Union (ERC, WANDA, 101039452). Views and opinions expressed are, however, those of the authors only and do not necessarily reflect those of the European Union or the European Research Council. Neither the European Union nor the granting authority can be held responsible for them. T.P.C. was supported by the Heising-Simons Foundation through a 51 Pegasi b Fellowship. Support for C.J.L. was provided by NASA through the NASA Hubble Fellowship grant No. HST-HF2-51535.001-A. This paper acknowledges the usage of ALMA data (project codes 2011.0.00526.S, 2012.1.00688.S, 2013.1.00395.S, 2018.1.00564.S), and VLT/X-Shooter spectra (project codes 097.C-0378, 0101.C-0866, 105.2082.003, 113.26NN.001, 113.26NN.003, 115.27XL.001).
\end{acknowledgements}

\vspace{-1cm}
\bibliographystyle{aa}
\bibliography{main}

@ARTICLE{Andrews_2013,
       author = {{Andrews}, Sean M. and {Rosenfeld}, Katherine A. and {Kraus}, Adam L. and {Wilner}, David J.},
        title = "{The Mass Dependence between Protoplanetary Disks and their Stellar Hosts}",
      journal = {\apj},
         year = 2013,
        month = jul,
       volume = {771},
       number = {2},
          eid = {129},
        pages = {129},
          doi = {10.1088/0004-637X/771/2/129},
archivePrefix = {arXiv},
       eprint = {1305.5262},
 primaryClass = {astro-ph.SR},
       adsurl = {https://ui.adsabs.harvard.edu/abs/2013ApJ...771..129A}
}

@ARTICLE{Zallio_2026,
       author = {{Zallio}, Luigi and {Rosotti}, Giovanni P. and {Vioque}, Miguel and {Miotello}, Anna and {Andrews}, Sean M. and {Manara}, Carlo F. and {Carpenter}, John M. and {Empey}, Aaron and {Kurtovic}, Nicol{\'a}s T. and {Law}, Charles J. and {Longarini}, Cristiano and {Paneque-Carre{\~n}o}, Teresa and {Teague}, Richard and {Villenave}, Marion and {Yen}, Hsi-Wei and {Zagaria}, Francesco},
        title = "{The $^{12}$CO gas structures of protoplanetary disks in the Upper Scorpius region}",
      journal = {\aap},
         year = 2026,
        month = jan,
       volume = {705},
          eid = {A49},
        pages = {A49},
          doi = {10.1051/0004-6361/202557366},
archivePrefix = {arXiv},
       eprint = {2511.16734},
 primaryClass = {astro-ph.SR},
       adsurl = {https://ui.adsabs.harvard.edu/abs/2026A&A...705A..49Z}
}

@BOOK{Hartmann_1998,
       author = {{Hartmann}, Lee},
        title = "{Accretion Processes in Star Formation}",
         year = 1998,
       volume = {32},
       adsurl = {https://ui.adsabs.harvard.edu/abs/1998apsf.book.....H}
}

@ARTICLE{Pinte_2009,
       author = {{Pinte}, C. and {Harries}, T.~J. and {Min}, M. and {Watson}, A.~M. and {Dullemond}, C.~P. and {Woitke}, P. and {M{\'e}nard}, F. and {Dur{\'a}n-Rojas}, M.~C.},
        title = "{Benchmark problems for continuum radiative transfer. High optical depths, anisotropic scattering, and polarisation}",
      journal = {\aap},
         year = 2009,
        month = may,
       volume = {498},
       number = {3},
        pages = {967-980},
          doi = {10.1051/0004-6361/200811555},
archivePrefix = {arXiv},
       eprint = {0903.1231},
 primaryClass = {astro-ph.SR},
       adsurl = {https://ui.adsabs.harvard.edu/abs/2009A&A...498..967P}
}

@ARTICLE{Guilloteau_2014,
       author = {{Guilloteau}, S. and {Simon}, M. and {Pi{\'e}tu}, V. and {Di Folco}, E. and {Dutrey}, A. and {Prato}, L. and {Chapillon}, E.},
        title = "{The masses of young stars: CN as a probe of dynamical masses}",
      journal = {\aap},
         year = 2014,
        month = jul,
       volume = {567},
          eid = {A117},
        pages = {A117},
          doi = {10.1051/0004-6361/201423765},
archivePrefix = {arXiv},
       eprint = {1406.3805},
 primaryClass = {astro-ph.SR},
       adsurl = {https://ui.adsabs.harvard.edu/abs/2014A&A...567A.117G}
}

@ARTICLE{Manara_2013,
       author = {{Manara}, C.~F. and {Testi}, L. and {Rigliaco}, E. and {Alcal{\'a}}, J.~M. and {Natta}, A. and {Stelzer}, B. and {Biazzo}, K. and {Covino}, E. and {Covino}, S. and {Cupani}, G. and {D'Elia}, V. and {Randich}, S.},
        title = "{X-shooter spectroscopy of young stellar objects. II. Impact of chromospheric emission on accretion rate estimates}",
      journal = {\aap},
         year = 2013,
        month = mar,
       volume = {551},
          eid = {A107},
        pages = {A107},
          doi = {10.1051/0004-6361/201220921},
archivePrefix = {arXiv},
       eprint = {1301.3058},
 primaryClass = {astro-ph.GA},
       adsurl = {https://ui.adsabs.harvard.edu/abs/2013A&A...551A.107M}
}

@ARTICLE{Herczeg_2014,
       author = {{Herczeg}, Gregory J. and {Hillenbrand}, Lynne A.},
        title = "{An Optical Spectroscopic Study of T Tauri Stars. I. Photospheric Properties}",
      journal = {\apj},
         year = 2014,
        month = may,
       volume = {786},
       number = {2},
          eid = {97},
        pages = {97},
          doi = {10.1088/0004-637X/786/2/97},
archivePrefix = {arXiv},
       eprint = {1403.1675},
 primaryClass = {astro-ph.SR},
       adsurl = {https://ui.adsabs.harvard.edu/abs/2014ApJ...786...97H}
}

@ARTICLE{Ansdell_2017,
       author = {{Ansdell}, M. and {Williams}, J.~P. and {Manara}, C.~F. and {Miotello}, A. and {Facchini}, S. and {van der Marel}, N. and {Testi}, L. and {van Dishoeck}, E.~F.},
        title = "{An ALMA Survey of Protoplanetary Disks in the {\ensuremath{\sigma}} Orionis Cluster}",
      journal = {\aj},
         year = 2017,
        month = may,
       volume = {153},
       number = {5},
          eid = {240},
        pages = {240},
          doi = {10.3847/1538-3881/aa69c0},
archivePrefix = {arXiv},
       eprint = {1703.08546},
 primaryClass = {astro-ph.EP},
       adsurl = {https://ui.adsabs.harvard.edu/abs/2017AJ....153..240A}
}

@article{Waskom2021,
    doi = {10.21105/joss.03021},
    url = {https://doi.org/10.21105/joss.03021},
    year = {2021},
    publisher = {The Open Journal},
    volume = {6},
    number = {60},
    pages = {3021},
    author = {Michael L. Waskom},
    title = {seaborn: statistical data visualization},
    journal = {Journal of Open Source Software}
 }

@ARTICLE{Rosenfeld_2012,
       author = {{Rosenfeld}, Katherine A. and {Andrews}, Sean M. and {Wilner}, David J. and {Stempels}, H.~C.},
        title = "{A Disk-based Dynamical Mass Estimate for the Young Binary V4046 Sgr}",
      journal = {\apj},
         year = 2012,
        month = nov,
       volume = {759},
       number = {2},
          eid = {119},
        pages = {119},
          doi = {10.1088/0004-637X/759/2/119},
archivePrefix = {arXiv},
       eprint = {1209.4407},
 primaryClass = {astro-ph.SR},
       adsurl = {https://ui.adsabs.harvard.edu/abs/2012ApJ...759..119R}
}

@ARTICLE{Gangi_2022,
       author = {{Gangi}, M. and {Antoniucci}, S. and {Biazzo}, K. and {Frasca}, A. and {Nisini}, B. and {Alcal{\'a}}, J.~M. and {Giannini}, T. and {Manara}, C.~F. and {Giunta}, A. and {Harutyunyan}, A. and {Munari}, U. and {Vitali}, F.},
        title = "{GIARPS High-resolution Observations of T Tauri stars (GHOsT). IV. Accretion properties of the Taurus-Auriga young association}",
      journal = {\aap},
         year = 2022,
        month = nov,
       volume = {667},
          eid = {A124},
        pages = {A124},
          doi = {10.1051/0004-6361/202244042},
archivePrefix = {arXiv},
       eprint = {2208.14895},
 primaryClass = {astro-ph.SR},
       adsurl = {https://ui.adsabs.harvard.edu/abs/2022A&A...667A.124G}
}

@ARTICLE{Towner_2025,
       author = {{Towner}, A.~P.~M. and {Eisner}, J.~A. and {Sheehan}, P.~D. and {Hillenbrand}, L.~A. and {Wu}, Y.-L.},
        title = "{Dynamical Masses for 23 Pre-main-sequence Stars in Upper Scorpius: A Critical Test of Stellar Evolutionary Models}",
      journal = {\apj},
         year = 2025,
        month = dec,
       volume = {994},
       number = {2},
          eid = {214},
        pages = {214},
          doi = {10.3847/1538-4357/ae0f18},
archivePrefix = {arXiv},
       eprint = {2509.23001},
 primaryClass = {astro-ph.GA},
       adsurl = {https://ui.adsabs.harvard.edu/abs/2025ApJ...994..214T}
}

@ARTICLE{Fang_2025,
       author = {{Fang}, Min and {Herczeg}, Gregory J.},
        title = "{A Comprehensive Gaia Spectroscopic Study of Stars in the Scorpius-Centaurus Complex: Star Formation History and Disk Lifetime}",
      journal = {\apj},
         year = 2025,
        month = dec,
       volume = {994},
       number = {2},
          eid = {248},
        pages = {248},
          doi = {10.3847/1538-4357/ae0f9a},
archivePrefix = {arXiv},
       eprint = {2509.19693},
 primaryClass = {astro-ph.SR},
       adsurl = {https://ui.adsabs.harvard.edu/abs/2025ApJ...994..248F}
}

@ARTICLE{2016ARA&A..54..135H,
       author = {{Hartmann}, Lee and {Herczeg}, Gregory and {Calvet}, Nuria},
        title = "{Accretion onto Pre-Main-Sequence Stars}",
      journal = {\araa},
         year = 2016,
        month = sep,
       volume = {54},
        pages = {135-180},
          doi = {10.1146/annurev-astro-081915-023347},
       adsurl = {https://ui.adsabs.harvard.edu/abs/2016ARA&A..54..135H}
}

@ARTICLE{2003ApJ...592..266M,
       author = {{Muzerolle}, James and {Hillenbrand}, Lynne and {Calvet}, Nuria and {Brice{\~n}o}, C{\'e}sar and {Hartmann}, Lee},
        title = "{Accretion in Young Stellar/Substellar Objects}",
      journal = {\apj},
         year = 2003,
        month = jul,
       volume = {592},
       number = {1},
        pages = {266-281},
          doi = {10.1086/375704},
archivePrefix = {arXiv},
       eprint = {astro-ph/0304078},
 primaryClass = {astro-ph},
       adsurl = {https://ui.adsabs.harvard.edu/abs/2003ApJ...592..266M}
}

@ARTICLE{2022ApJ...930...39V,
       author = {{Vioque}, Miguel and {Oudmaijer}, Ren{\'e} D. and {Wichittanakom}, Chumpon and {Mendigut{\'\i}a}, Ignacio and {Baines}, Deborah and {Pani{\'c}}, Olja and {Iglesias}, Daniela and {Miley}, James and {P{\'e}rez-Mart{\'\i}nez}, Ricardo},
        title = "{Identification and Spectroscopic Characterization of 128 New Herbig Stars}",
      journal = {\apj},
         year = 2022,
        month = may,
       volume = {930},
       number = {1},
          eid = {39},
        pages = {39},
          doi = {10.3847/1538-4357/ac5c46},
archivePrefix = {arXiv},
       eprint = {2202.01234},
 primaryClass = {astro-ph.SR},
       adsurl = {https://ui.adsabs.harvard.edu/abs/2022ApJ...930...39V}
}

@ARTICLE{2022A&A...659A..85F,
       author = {{Franciosini}, E. and {Tognelli}, E. and {Degl'Innocenti}, S. and {Prada Moroni}, P.~G. and {Randich}, S. and {Sacco}, G.~G. and {Magrini}, L. and {Pancino}, E. and {Lanzafame}, A.~C. and {Smiljanic}, R. and {Prisinzano}, L. and {Sanna}, N. and {Roccatagliata}, V. and {Bonito}, R. and {de Laverny}, P. and {Guti{\'e}rrez Albarr{\'a}n}, M.~L. and {Montes}, D. and {Jim{\'e}nez-Esteban}, F. and {Gilmore}, G. and {Bergemann}, M. and {Carraro}, G. and {Damiani}, F. and {Gonneau}, A. and {Hourihane}, A. and {Morbidelli}, L. and {Worley}, C.~C. and {Zaggia}, S.},
        title = "{Gaia-ESO Survey: Role of magnetic activity and starspots on pre-main-sequence lithium evolution}",
      journal = {\aap},
         year = 2022,
        month = mar,
       volume = {659},
          eid = {A85},
        pages = {A85},
          doi = {10.1051/0004-6361/202142290},
archivePrefix = {arXiv},
       eprint = {2111.11196},
 primaryClass = {astro-ph.SR},
       adsurl = {https://ui.adsabs.harvard.edu/abs/2022A&A...659A..85F}
}

@ARTICLE{2024NatAs...8..216M,
       author = {{Miret-Roig}, N{\'u}ria and {Alves}, Jo{\~a}o and {Barrado}, David and {Burkert}, Andreas and {Ratzenb{\"o}ck}, Sebastian and {Konietzka}, Ralf},
        title = "{Insights into star formation and dispersal from the synchronization of stellar clocks}",
      journal = {Nature Astronomy},
         year = 2024,
        month = feb,
       volume = {8},
        pages = {216-222},
          doi = {10.1038/s41550-023-02132-4},
archivePrefix = {arXiv},
       eprint = {2311.13042},
 primaryClass = {astro-ph.SR},
       adsurl = {https://ui.adsabs.harvard.edu/abs/2024NatAs...8..216M}
}

@ARTICLE{2011ApJ...738..140H,
       author = {{Hosokawa}, Takashi and {Offner}, Stella S.~R. and {Krumholz}, Mark R.},
        title = "{On the Reliability of Stellar Ages and Age Spreads Inferred from Pre-main-sequence Evolutionary Models}",
      journal = {\apj},
         year = 2011,
        month = sep,
       volume = {738},
       number = {2},
          eid = {140},
        pages = {140},
          doi = {10.1088/0004-637X/738/2/140},
archivePrefix = {arXiv},
       eprint = {1101.3599},
 primaryClass = {astro-ph.SR},
       adsurl = {https://ui.adsabs.harvard.edu/abs/2011ApJ...738..140H}
}

@INPROCEEDINGS{2014prpl.conf..219S,
       author = {{Soderblom}, D.~R. and {Hillenbrand}, L.~A. and {Jeffries}, R.~D. and {Mamajek}, E.~E. and {Naylor}, T.},
        title = "{Ages of Young Stars}",
    booktitle = {Protostars and Planets VI},
         year = 2014,
       editor = {{Beuther}, Henrik and {Klessen}, Ralf S. and {Dullemond}, Cornelis P. and {Henning}, Thomas},
        month = jan,
        pages = {219-241},
          doi = {10.2458/azu_uapress_9780816531240-ch010},
archivePrefix = {arXiv},
       eprint = {1311.7024},
 primaryClass = {astro-ph.SR},
       adsurl = {https://ui.adsabs.harvard.edu/abs/2014prpl.conf..219S}
}

@ARTICLE{2012ApJ...746..154P,
       author = {{Pecaut}, Mark J. and {Mamajek}, Eric E. and {Bubar}, Eric J.},
        title = "{A Revised Age for Upper Scorpius and the Star Formation History among the F-type Members of the Scorpius-Centaurus OB Association}",
      journal = {\apj},
         year = 2012,
        month = feb,
       volume = {746},
       number = {2},
          eid = {154},
        pages = {154},
          doi = {10.1088/0004-637X/746/2/154},
archivePrefix = {arXiv},
       eprint = {1112.1695},
 primaryClass = {astro-ph.SR},
       adsurl = {https://ui.adsabs.harvard.edu/abs/2012ApJ...746..154P}
}

@INPROCEEDINGS{2008ASPC..384..200H,
       author = {{Hillenbrand}, L.~A. and {Bauermeister}, A. and {White}, R.~J.},
        title = "{An Assessment of HR Diagram Constraints on Ages and Age Spreads in Star-Forming Regions and Young Clusters}",
    booktitle = {14th Cambridge Workshop on Cool Stars, Stellar Systems, and the Sun},
         year = 2008,
       editor = {{van Belle}, G.},
       series = {Astronomical Society of the Pacific Conference Series},
       volume = {384},
        month = apr,
        pages = {200},
          doi = {10.48550/arXiv.astro-ph/0703642},
archivePrefix = {arXiv},
       eprint = {astro-ph/0703642},
 primaryClass = {astro-ph},
       adsurl = {https://ui.adsabs.harvard.edu/abs/2008ASPC..384..200H}
}

@ARTICLE{2000ApJ...540..255P,
       author = {{Palla}, Francesco and {Stahler}, Steven W.},
        title = "{Accelerating Star Formation in Clusters and Associations}",
      journal = {\apj},
         year = 2000,
        month = sep,
       volume = {540},
       number = {1},
        pages = {255-270},
          doi = {10.1086/309312},
       adsurl = {https://ui.adsabs.harvard.edu/abs/2000ApJ...540..255P}
}

@ARTICLE{1965ApJ...142..841H,
       author = {{Henyey}, Louis and {Vardya}, M.~S. and {Bodenheimer}, Peter},
        title = "{Studies in Stellar Evolution. III. The Calculation of Model Envelopes.}",
      journal = {\apj},
         year = 1965,
        month = oct,
       volume = {142},
        pages = {841},
          doi = {10.1086/148357},
       adsurl = {https://ui.adsabs.harvard.edu/abs/1965ApJ...142..841H}
}

@ARTICLE{1961PASJ...13..450H,
       author = {{Hayashi}, Chushiro},
        title = "{Stellar evolution in early phases of gravitational contraction.}",
      journal = {\pasj},
         year = 1961,
        month = dec,
       volume = {13},
        pages = {450-452},
       adsurl = {https://ui.adsabs.harvard.edu/abs/1961PASJ...13..450H}
}

@ARTICLE{Gully-Santiago_2017,
       author = {{Gully-Santiago}, Michael A. and {Herczeg}, Gregory J. and {Czekala}, Ian and {Somers}, Garrett and {Grankin}, Konstantin and {Covey}, Kevin R. and {Donati}, J.~F. and {Alencar}, Silvia H.~P. and {Hussain}, Gaitee A.~J. and {Shappee}, Benjamin J. and {Mace}, Gregory N. and {Lee}, Jae-Joon and {Holoien}, T.~W. -S. and {Jose}, Jessy and {Liu}, Chun-Fan},
        title = "{Placing the Spotted T Tauri Star LkCa 4 on an HR Diagram}",
      journal = {\apj},
         year = 2017,
        month = feb,
       volume = {836},
       number = {2},
          eid = {200},
        pages = {200},
          doi = {10.3847/1538-4357/836/2/200},
archivePrefix = {arXiv},
       eprint = {1701.06703},
 primaryClass = {astro-ph.SR},
       adsurl = {https://ui.adsabs.harvard.edu/abs/2017ApJ...836..200G}
}

@ARTICLE{Grankin_2008,
       author = {{Grankin}, K.~N. and {Bouvier}, J. and {Herbst}, W. and {Melnikov}, S. Yu.},
        title = "{Results of the ROTOR-program. II. The long-term photometric variability of weak-line T Tauri stars}",
      journal = {\aap},
         year = 2008,
        month = mar,
       volume = {479},
       number = {3},
        pages = {827-843},
          doi = {10.1051/0004-6361:20078476},
archivePrefix = {arXiv},
       eprint = {0801.3543},
 primaryClass = {astro-ph},
       adsurl = {https://ui.adsabs.harvard.edu/abs/2008A&A...479..827G}
}

@ARTICLE{Facundo_2023,
       author = {{P{\'e}rez Paolino}, Facundo and {Bary}, Jeffrey S. and {Petersen}, Michael S. and {Ward-Duong}, Kimberly and {Tofflemire}, Benjamin M. and {Follette}, Katherine B. and {Mach}, Heidi},
        title = "{Correlating Changes in Spot Filling Factors with Stellar Rotation: The Case of LkCa 4}",
      journal = {\apj},
         year = 2023,
        month = mar,
       volume = {946},
       number = {1},
          eid = {10},
        pages = {10},
          doi = {10.3847/1538-4357/acbb61},
archivePrefix = {arXiv},
       eprint = {2303.01574},
 primaryClass = {astro-ph.SR},
       adsurl = {https://ui.adsabs.harvard.edu/abs/2023ApJ...946...10P}
}

@article{Pascucci_2016,
author = {{Pascucci}, I. and {Testi}, L. and {Herczeg}, G.~J. and {Long}, F. and {Manara}, C.~F. and {Hendler}, N. and {Mulders}, G.~D. and {Krijt}, S. and {Ciesla}, F. and {Henning}, Th. and {Mohanty}, S. and {Drabek-Maunder}, E. and {Apai}, D. and {Sz{\H{u}}cs}, L. and {Sacco}, G. and {Olofsson}, J.},
title = {A Steeper than Linear Disk Mass-Stellar Mass Scaling Relation},
journal = {The Astrophysical Journal},
year = 2016,
month = nov,
volume = {831},
number = {2},
eid = {125},
pages = {125},
doi = {10.3847/0004-637X/831/2/125},
archivePrefix = {arXiv},
eprint = {1608.03621},
primaryClass = {astro-ph.EP},
adsurl = {https://ui.adsabs.harvard.edu/abs/2016ApJ...831..125P}
}

@ARTICLE{Facundo_2024,
       author = {{P{\'e}rez Paolino}, Facundo and {Bary}, Jeffrey S. and {Hillenbrand}, Lynne A. and {Markham}, Madison},
        title = "{The Effect of Starspots on Spectroscopic Age and Mass Estimates of Nonaccreting T Tauri Stars in the Taurus{\textendash}Auriga Star-forming Region}",
      journal = {\apj},
         year = 2024,
        month = may,
       volume = {967},
       number = {1},
          eid = {45},
        pages = {45},
          doi = {10.3847/1538-4357/ad393b},
archivePrefix = {arXiv},
       eprint = {2403.20255},
 primaryClass = {astro-ph.SR},
       adsurl = {https://ui.adsabs.harvard.edu/abs/2024ApJ...967...45P}
}

@ARTICLE{Simon_2000,
       author = {{Simon}, M. and {Dutrey}, A. and {Guilloteau}, S.},
        title = "{Dynamical Masses of T Tauri Stars and Calibration of Pre-Main-Sequence Evolution}",
      journal = {\apj},
         year = 2000,
        month = dec,
       volume = {545},
       number = {2},
        pages = {1034-1043},
          doi = {10.1086/317838},
archivePrefix = {arXiv},
       eprint = {astro-ph/0008370},
 primaryClass = {astro-ph},
       adsurl = {https://ui.adsabs.harvard.edu/abs/2000ApJ...545.1034S}
}

@ARTICLE{Braun_2021,
       author = {{Braun}, Teresa A.~M. and {Yen}, Hsi-Wei and {Koch}, Patrick M. and {Manara}, Carlo F. and {Miotello}, Anna and {Testi}, Leonardo},
        title = "{Dynamical Stellar Masses of Pre-main-sequence Stars in Lupus and Taurus Obtained with ALMA Surveys in Comparison with Stellar Evolutionary Models}",
      journal = {\apj},
         year = 2021,
        month = feb,
       volume = {908},
       number = {1},
          eid = {46},
        pages = {46},
          doi = {10.3847/1538-4357/abd24f},
archivePrefix = {arXiv},
       eprint = {2012.07441},
 primaryClass = {astro-ph.SR},
       adsurl = {https://ui.adsabs.harvard.edu/abs/2021ApJ...908...46B}
}

@ARTICLE{Andrews_2024,
       author = {{Andrews}, Sean M. and {Teague}, Richard and {Wirth}, Christopher P. and {Huang}, Jane and {Zhu}, Zhaohuan},
        title = "{On Kinematic Measurements of Self-gravity in Protoplanetary Disks}",
      journal = {\apj},
         year = 2024,
        month = aug,
       volume = {970},
       number = {2},
          eid = {153},
        pages = {153},
          doi = {10.3847/1538-4357/ad5285},
archivePrefix = {arXiv},
       eprint = {2405.19574},
 primaryClass = {astro-ph.EP},
       adsurl = {https://ui.adsabs.harvard.edu/abs/2024ApJ...970..153A}
}

@ARTICLE{Somers_2020,
       author = {{Somers}, Garrett and {Cao}, Lyra and {Pinsonneault}, Marc H.},
        title = "{The SPOTS Models: A Grid of Theoretical Stellar Evolution Tracks and Isochrones for Testing the Effects of Starspots on Structure and Colors}",
      journal = {\apj},
         year = 2020,
        month = mar,
       volume = {891},
       number = {1},
          eid = {29},
        pages = {29},
          doi = {10.3847/1538-4357/ab722e},
archivePrefix = {arXiv},
       eprint = {2002.10644},
 primaryClass = {astro-ph.SR},
       adsurl = {https://ui.adsabs.harvard.edu/abs/2020ApJ...891...29S}
}

@ARTICLE{Nguyen_2022,
       author = {{Nguyen}, C.~T. and {Costa}, G. and {Girardi}, L. and {Volpato}, G. and {Bressan}, A. and {Chen}, Y. and {Marigo}, P. and {Fu}, X. and {Goudfrooij}, P.},
        title = "{PARSEC V2.0: Stellar tracks and isochrones of low- and intermediate-mass stars with rotation}",
      journal = {\aap},
         year = 2022,
        month = sep,
       volume = {665},
          eid = {A126},
        pages = {A126},
          doi = {10.1051/0004-6361/202244166},
archivePrefix = {arXiv},
       eprint = {2207.08642},
 primaryClass = {astro-ph.SR},
       adsurl = {https://ui.adsabs.harvard.edu/abs/2022A&A...665A.126N}
}

@ARTICLE{Ratzenbock_2023a,
       author = {{Ratzenb{\"o}ck}, Sebastian and {Gro{\ss}schedl}, Josefa E. and {Alves}, Jo{\~a}o and {Miret-Roig}, N{\'u}ria and {Bomze}, Immanuel and {Forbes}, John and {Goodman}, Alyssa and {Hacar}, {\'A}lvaro and {Lin}, Doug and {Meingast}, Stefan and {M{\"o}ller}, Torsten and {Piecka}, Martin and {Posch}, Laura and {Rottensteiner}, Alena and {Swiggum}, Cameren and {Zucker}, Catherine},
        title = "{The star formation history of the Sco-Cen association. Coherent star formation patterns in space and time}",
      journal = {\aap},
         year = 2023,
        month = oct,
       volume = {678},
          eid = {A71},
        pages = {A71},
          doi = {10.1051/0004-6361/202346901},
archivePrefix = {arXiv},
       eprint = {2302.07853},
 primaryClass = {astro-ph.SR},
       adsurl = {https://ui.adsabs.harvard.edu/abs/2023A&A...678A..71R}
}

@ARTICLE{Ratzenbock_2023b,
       author = {{Ratzenb{\"o}ck}, Sebastian and {Gro{\ss}schedl}, Josefa E. and {M{\"o}ller}, Torsten and {Alves}, Jo{\~a}o and {Bomze}, Immanuel and {Meingast}, Stefan},
        title = "{Significance mode analysis (SigMA) for hierarchical structures. An application to the Sco-Cen OB association}",
      journal = {\aap},
         year = 2023,
        month = sep,
       volume = {677},
          eid = {A59},
        pages = {A59},
          doi = {10.1051/0004-6361/202243690},
archivePrefix = {arXiv},
       eprint = {2211.14225},
 primaryClass = {astro-ph.GA},
       adsurl = {https://ui.adsabs.harvard.edu/abs/2023A&A...677A..59R}
}

@ARTICLE{Longarini_2025,
       author = {{Longarini}, Cristiano and {Lodato}, Giuseppe and {Rosotti}, Giovanni and {Andrews}, Sean and {Winter}, Andrew and {Stadler}, Jochen and {Izquierdo}, Andr{\'e}s and {Galloway-Sprietsma}, Maria and {Facchini}, Stefano and {Curone}, Pietro and {Benisty}, Myriam and {Teague}, Richard and {Bae}, Jaehan and {Barraza-Alfaro}, Marcelo and {Cataldi}, Gianni and {Czekala}, Ian and {Cuello}, Nicol{\'a}s and {Fasano}, Daniele and {Flock}, Mario and {Fukagawa}, Misato and {Garg}, Himanshi and {Hall}, Cassandra and {Hammond}, Iain and {Hardiman}, Caitlyn and {Hilder}, Thomas and {Huang}, Jane and {Ilee}, John D. and {Isella}, Andrea and {Kanagawa}, Kazuhiro and {Lesur}, Geoffroy and {Loomis}, Ryan A. and {M{\'e}nard}, Francois and {Orihara}, Ryuta and {Pinte}, Christophe and {Price}, Daniel and {Testi}, Leonardo and {Fernandez}, Gaylor Wafflard- and {W{\"o}lfer}, Lisa and {Yen}, Hsi-Wei and {Yoshida}, Tomohiro C. and {Zawadzki}, Brianna},
        title = "{exoALMA. XII. Weighing and Sizing exoALMA Disks with Rotation Curve Modelling}",
      journal = {\apjl},
         year = 2025,
        month = may,
       volume = {984},
       number = {1},
          eid = {L17},
        pages = {L17},
          doi = {10.3847/2041-8213/adc431},
archivePrefix = {arXiv},
       eprint = {2504.18726},
 primaryClass = {astro-ph.EP},
       adsurl = {https://ui.adsabs.harvard.edu/abs/2025ApJ...984L..17L}
}

@ARTICLE{Simon_2019,
       author = {{Simon}, M. and {Guilloteau}, S. and {Beck}, Tracy L. and {Chapillon}, E. and {Di Folco}, E. and {Dutrey}, A. and {Feiden}, Gregory A. and {Grosso}, N. and {Pi{\'e}tu}, V. and {Prato}, L. and {Schaefer}, Gail H.},
        title = "{Masses and Implications for Ages of Low-mass Pre-main-sequence Stars in Taurus and Ophiuchus}",
      journal = {\apj},
         year = 2019,
        month = oct,
       volume = {884},
       number = {1},
          eid = {42},
        pages = {42},
          doi = {10.3847/1538-4357/ab3e3b},
archivePrefix = {arXiv},
       eprint = {1908.10952},
 primaryClass = {astro-ph.SR},
       adsurl = {https://ui.adsabs.harvard.edu/abs/2019ApJ...884...42S}
}

@ARTICLE{Simon_2017,
       author = {{Simon}, M. and {Guilloteau}, S. and {Di Folco}, E. and {Dutrey}, A. and {Grosso}, N. and {Pi{\'e}tu}, V. and {Chapillon}, E. and {Prato}, L. and {Schaefer}, G.~H. and {Rice}, E. and {Boehler}, Y.},
        title = "{Dynamical Masses of Low-mass Stars in the Taurus and Ophiuchus Star-forming Regions}",
      journal = {\apj},
         year = 2017,
        month = aug,
       volume = {844},
       number = {2},
          eid = {158},
        pages = {158},
          doi = {10.3847/1538-4357/aa78f1},
archivePrefix = {arXiv},
       eprint = {1706.03505},
 primaryClass = {astro-ph.SR},
       adsurl = {https://ui.adsabs.harvard.edu/abs/2017ApJ...844..158S}
}

@ARTICLE{Yen_2018,
       author = {{Yen}, Hsi-Wei and {Koch}, Patrick M. and {Manara}, Carlo F. and {Miotello}, Anna and {Testi}, Leonardo},
        title = "{Stellar masses and disk properties of Lupus young stellar objects traced by velocity-aligned stacked ALMA $^{13}$CO and C$^{18}$O spectra}",
      journal = {\aap},
         year = 2018,
        month = aug,
       volume = {616},
          eid = {A100},
        pages = {A100},
          doi = {10.1051/0004-6361/201732196},
archivePrefix = {arXiv},
       eprint = {1804.06272},
 primaryClass = {astro-ph.GA},
       adsurl = {https://ui.adsabs.harvard.edu/abs/2018A&A...616A.100Y}
}

@ARTICLE{Czekala_2016,
       author = {{Czekala}, I. and {Andrews}, S.~M. and {Torres}, G. and {Jensen}, E.~L.~N. and {Stassun}, K.~G. and {Wilner}, D.~J. and {Latham}, D.~W.},
        title = "{A Disk-based Dynamical Constraint on the Mass of the Young Binary DQ Tau}",
      journal = {\apj},
         year = 2016,
        month = feb,
       volume = {818},
       number = {2},
          eid = {156},
        pages = {156},
          doi = {10.3847/0004-637X/818/2/156},
archivePrefix = {arXiv},
       eprint = {1601.03806},
 primaryClass = {astro-ph.SR},
       adsurl = {https://ui.adsabs.harvard.edu/abs/2016ApJ...818..156C}
}

@ARTICLE{Sheehan_2019,
       author = {{Sheehan}, Patrick D. and {Wu}, Ya-Lin and {Eisner}, Josh A. and {Tobin}, John J.},
        title = "{High-precision Dynamical Masses of Pre-main-sequence Stars with ALMA and Gaia}",
      journal = {\apj},
         year = 2019,
        month = apr,
       volume = {874},
       number = {2},
          eid = {136},
        pages = {136},
          doi = {10.3847/1538-4357/ab09f9},
archivePrefix = {arXiv},
       eprint = {1903.00032},
 primaryClass = {astro-ph.SR},
       adsurl = {https://ui.adsabs.harvard.edu/abs/2019ApJ...874..136S}
}

@ARTICLE{Siess_2000,
       author = {{Siess}, L. and {Dufour}, E. and {Forestini}, M.},
        title = "{An internet server for pre-main sequence tracks of low- and intermediate-mass stars}",
      journal = {\aap},
         year = 2000,
        month = jun,
       volume = {358},
        pages = {593-599},
          doi = {10.48550/arXiv.astro-ph/0003477},
archivePrefix = {arXiv},
       eprint = {astro-ph/0003477},
 primaryClass = {astro-ph},
       adsurl = {https://ui.adsabs.harvard.edu/abs/2000A&A...358..593S}
}

@ARTICLE{Pinte_2006,
       author = {{Pinte}, C. and {M{\'e}nard}, F. and {Duch{\^e}ne}, G. and {Bastien}, P.},
        title = "{Monte Carlo radiative transfer in protoplanetary disks}",
      journal = {\aap},
         year = 2006,
        month = dec,
       volume = {459},
       number = {3},
        pages = {797-804},
          doi = {10.1051/0004-6361:20053275},
archivePrefix = {arXiv},
       eprint = {astro-ph/0606550},
 primaryClass = {astro-ph},
       adsurl = {https://ui.adsabs.harvard.edu/abs/2006A&A...459..797P}
}

@ARTICLE{Claes_2024,
       author = {{Claes}, R.~A.~B. and {Campbell-White}, J. and {Manara}, C.~F. and {Frasca}, A. and {Natta}, A. and {Alcal{\'a}}, J.~M. and {Armeni}, A. and {Fang}, M. and {Lovell}, J.~B. and {Stelzer}, B. and {Venuti}, L. and {Wyatt}, M. and {Queitsch}, A.},
        title = "{FitteR for Accretion ProPErties of T Tauri stars (FRAPPE): A new approach to use class III spectra to derive stellar and accretion properties}",
      journal = {\aap},
         year = 2024,
        month = oct,
       volume = {690},
          eid = {A122},
        pages = {A122},
          doi = {10.1051/0004-6361/202450885},
archivePrefix = {arXiv},
       eprint = {2407.11866},
 primaryClass = {astro-ph.SR},
       adsurl = {https://ui.adsabs.harvard.edu/abs/2024A&A...690A.122C}
}

@ARTICLE{Foreman_2013,
       author = {{Foreman-Mackey}, Daniel and {Hogg}, David W. and {Lang}, Dustin and {Goodman}, Jonathan},
        title = "{emcee: The MCMC Hammer}",
      journal = {\pasp},
         year = 2013,
        month = mar,
       volume = {125},
       number = {925},
        pages = {306},
          doi = {10.1086/670067},
archivePrefix = {arXiv},
       eprint = {1202.3665},
 primaryClass = {astro-ph.IM},
       adsurl = {https://ui.adsabs.harvard.edu/abs/2013PASP..125..306F}
}

@ARTICLE{Deng_2025,
       author = {{Deng}, Dingshan and {Vioque}, Miguel and {Pascucci}, Ilaria and {P{\'e}rez}, Laura M. and {Zhang}, Ke and {Kurtovic}, Nicol{\'a}s and {Trapman}, Leon and {TorresVillanueva}, Estephani E. and {Agurto-Gangas}, Carolina and {Carpenter}, John and {Pinilla}, Paola and {Gorti}, Uma and {Tabone}, Beno{\^\i}t and {Sierra}, Anibal and {Rosotti}, Giovanni P. and {Cieza}, Lucas A. and {Anania}, Rossella and {Gonz{\'a}lez-Ruilova}, Camilo and {Hogerheijde}, Michiel R. and {Miley}, James and {Ruiz-Rodriguez}, Dary A. and {Ruaud}, Maxime and {Schwarz}, Kamber},
        title = "{The ALMA Survey of Gas Evolution of PROtoplanetary Disks (AGE-PRO). III. Dust and Gas Disk Properties in the Lupus Star-forming Region}",
      journal = {\apj},
         year = 2025,
        month = aug,
       volume = {989},
       number = {1},
          eid = {3},
        pages = {3},
          doi = {10.3847/1538-4357/add43a},
archivePrefix = {arXiv},
       eprint = {2506.10734},
 primaryClass = {astro-ph.EP},
       adsurl = {https://ui.adsabs.harvard.edu/abs/2025ApJ...989....3D}
}

@ARTICLE{Feiden2016,
       author = {{Feiden}, Gregory A.},
        title = "{Magnetic inhibition of convection and the fundamental properties of low-mass stars. III. A consistent 10 Myr age for the Upper Scorpius OB association}",
      journal = {\aap},
         year = 2016,
        month = sep,
       volume = {593},
          eid = {A99},
        pages = {A99},
          doi = {10.1051/0004-6361/201527613},
archivePrefix = {arXiv},
       eprint = {1604.08036},
 primaryClass = {astro-ph.SR},
       adsurl = {https://ui.adsabs.harvard.edu/abs/2016A&A...593A..99F}
}

@ARTICLE{Baraffe2015,
       author = {{Baraffe}, Isabelle and {Homeier}, Derek and {Allard}, France and {Chabrier}, Gilles},
        title = "{New evolutionary models for pre-main sequence and main sequence low-mass stars down to the hydrogen-burning limit}",
      journal = {\aap},
         year = 2015,
        month = may,
       volume = {577},
          eid = {A42},
        pages = {A42},
          doi = {10.1051/0004-6361/201425481},
archivePrefix = {arXiv},
       eprint = {1503.04107},
 primaryClass = {astro-ph.SR},
       adsurl = {https://ui.adsabs.harvard.edu/abs/2015A&A...577A..42B}
}

@INPROCEEDINGS{Manara_2023,
       author = {{Manara}, C.~F. and {Ansdell}, M. and {Rosotti}, G.~P. and {Hughes}, A.~M. and {Armitage}, P.~J. and {Lodato}, G. and {Williams}, J.~P.},
        title = "{Demographics of Young Stars and their Protoplanetary Disks: Lessons Learned on Disk Evolution and its Connection to Planet Formation}",
    booktitle = {Protostars and Planets VII},
         year = 2023,
       editor = {{Inutsuka}, S. and {Aikawa}, Y. and {Muto}, T. and {Tomida}, K. and {Tamura}, M.},
       series = {Astronomical Society of the Pacific Conference Series},
       volume = {534},
        month = jul,
        pages = {539},
          doi = {10.48550/arXiv.2203.09930},
archivePrefix = {arXiv},
       eprint = {2203.09930},
 primaryClass = {astro-ph.SR},
       adsurl = {https://ui.adsabs.harvard.edu/abs/2023ASPC..534..539M}
}

@ARTICLE{Manara_2020,
       author = {{Manara}, C.~F. and {Natta}, A. and {Rosotti}, G.~P. and {Alcal{\'a}}, J.~M. and {Nisini}, B. and {Lodato}, G. and {Testi}, L. and {Pascucci}, I. and {Hillenbrand}, L. and {Carpenter}, J. and {Scholz}, A. and {Fedele}, D. and {Frasca}, A. and {Mulders}, G. and {Rigliaco}, E. and {Scardoni}, C. and {Zari}, E.},
        title = "{X-shooter survey of disk accretion in Upper Scorpius. I. Very high accretion rates at age > 5 Myr}",
      journal = {\aap},
         year = 2020,
        month = jul,
       volume = {639},
          eid = {A58},
        pages = {A58},
          doi = {10.1051/0004-6361/202037949},
archivePrefix = {arXiv},
       eprint = {2004.14232},
 primaryClass = {astro-ph.SR},
       adsurl = {https://ui.adsabs.harvard.edu/abs/2020A&A...639A..58M}
}

@ARTICLE{Carpenter2025,
       author = {{Carpenter}, John M. and {Esplin}, Taran L. and {Luhman}, Kevin L. and {Mamajek}, Eric E. and {Andrews}, Sean M.},
        title = "{Extending the ALMA Census of Circumstellar Disks in the Upper Scorpius OB Association}",
      journal = {\apj},
         year = 2025,
        month = jan,
       volume = {978},
       number = {1},
          eid = {117},
        pages = {117},
          doi = {10.3847/1538-4357/ad8ebc},
archivePrefix = {arXiv},
       eprint = {2410.21598},
 primaryClass = {astro-ph.SR},
       adsurl = {https://ui.adsabs.harvard.edu/abs/2025ApJ...978..117C}
}

\newpage
\section*{Author Affiliations}
\begin{enumerate}
  \item \textbf{Dipartimento di Fisica ‘Aldo Pontremoli’}, Università degli Studi di Milano, via G. Celoria 16, I-20133 Milano, Italy.
  \item \textbf{European Southern Observatory}, Karl-Schwarzschild-Strasse 2, D-85748 Garching bei München, Germany.
  \item \textbf{Center for Astrophysics \textbar\ Harvard \& Smithsonian}, 60 Garden Street, Cambridge, MA 02138, USA.
  \item \textbf{University College Dublin (UCD)}, Department of Physics, Belfield, Dublin 4, Ireland.
  \item \textbf{Joint ALMA Observatory}, Avenida Alonso de Córdova 3107, Vitacura, Santiago, Chile.
  \item \textbf{Lunar and Planetary Laboratory, the University of Arizona}, Tucson, AZ 85721, USA.
  \item \textbf{Max Planck Institute for Extraterrestrial Physics}, Giessenbachstrasse 1, D-85748 Garching, Germany.
  \item \textbf{Max-Planck-Institut fur Astronomie (MPIA)}, Konigstuhl 17, 69117 Heidelberg, Germany.
  \item \textbf{Department of Astronomy, University of Virginia, Charlottesville}, VA 22904, USA.
  \item \textbf{Institute of Astronomy, University of Cambridge}, Madingley Road, Cambridge, CB3 0HA, UK.
  \item \textbf{Department of Astronomy, University of Michigan}, 1085 South University Avenue, Ann Arbor, MI 48109, USA.
  \item \textbf{Department of Earth, Atmospheric, and Planetary Sciences, Massachusetts Institute of Technology}, Cambridge, MA 02139, USA.
  \item \textbf{Univ. Grenoble Alpes, CNRS, IPAG}, F-38000 Grenoble, France.
  \item \textbf{Academia Sinica Institute of Astronomy and Astrophysics}, 11F of Astronomy-Mathematics Building, AS/NTU, No.1, Sec 4, Roosevelt Rd, Taipei 106216, Taiwan.
\end{enumerate}

\newpage

\begin{appendix}

\section{Uncertainties on the dynamical mass measurements due to inclination, stellar mass, and disk size}
\label{appendix:masses}
\begin{figure*}
    \centering
    \includegraphics[width=0.9\linewidth]{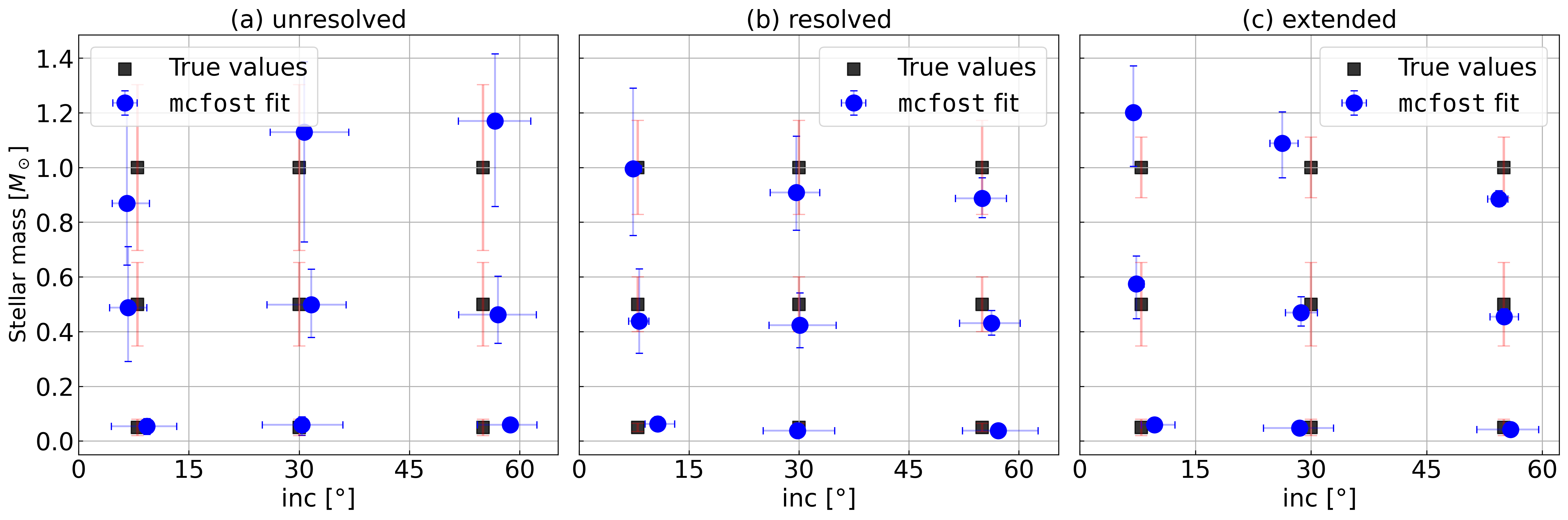}
    \caption{Results of the fit to unresolved (a), resolved (b), and extended (c) \texttt{mcfost} simulations. The uncertainties on top of the `True values' are the mean of the statistical uncertainties returned by the fit for each mass bin and for each disk size bin, and we plot them for illustrative purposes.}
    \label{fig:McFost_models}
\end{figure*}

Since the stellar mass statistical uncertainties reported by \cite{Zallio_2026} are systematically underestimated due to limitations of fitting visibilities with parametric models, we decide to test the accuracy of the \verb|csalt| dynamical mass measurements by creating and fitting a grid of numerical simulations. After creating \verb|mcfost|\footnote{\url{https://github.com/cpinte/mcfost}} (\citealt{Pinte_2006,Pinte_2009}) radiative transfer models of $^{12}$CO $J=3-2$, we use built-in routines of \verb|csalt| to transform them into radio-interferometric observed data, using the same total integration time ($\sim180$ s per disk), S/N, and antenna configurations of the observations analyzed in \cite{Zallio_2026}, from the sample of \citealt{Carpenter2025}).
Our grid of models tests the accuracy of mass inferences for stars with masses of $1$, $0.5$, and $0.05$ $M_{\odot}$ to explore stellar mass values down to the sub-stellar range, with disks inclined at 8°, 30°, and 55°. We perform these simulations using a gas disk mass of $10^{-2}\>M_\star$, a dust disk mass of $10^{-4}\>M_\star$, and three different cut-off radii ($R_{\rm{c}}=5, 20, 50$ au), mimicking unresolved (less than 3 resolution elements in the diameter), resolved (between 3 and 5 resolution elements in the diameter), and extended (more than 5 resolution elements in the disk diameter) protoplanetary disk extension. These three cases correspond to a disk size\footnote{We consider the radius that encompasses $90\%$ of the disk total flux $R_{90\%}$ as the gas disk radius.} $R_{90\%}<50$ au, $50\lesssim R_{90\%}\lesssim 90$ au, and $R_{90\%}>90$ au for the observations studied in \cite{Zallio_2026}. We then perform fits using \verb|csalt| with the parametric prescription A (see \citealt{Zallio_2026}; this is a complete disk prescription, which also considers for the emitting height of the optically thick $^{12}$CO layer) on the synthetic observation we created. Most of our fits return results in agreement with the true stellar mass, except for the sources with inc $=55$°, as shown in Fig. \ref{fig:McFost_models}, where the discrepancy is $\sim10\%$ in the worst case.

We use the statistical uncertainties returned by our fits per specific stellar mass range to assign mass uncertainties to the stellar dynamical masses reported in \cite{Zallio_2026}. We evaluate the mean statistical uncertainty for each stellar bin and for each disk size (reported in the third column of Table \ref{table:M_*_uncertainty}) of the uncertainties returned from the \verb|csalt| fits. This choice is motivated by the fact that the mean statistical uncertainties are found to be greater than the systematic uncertainty, and we therefore opt for a conservative approach. Then, we assign these uncertainties to our observations, as illustrated in Table \ref{table:M_*_uncertainty}, using the disk sizes reported in \cite{Zallio_2026}.
\begin{table}[t!]
    \centering
    \begin{tabular}{l c r}
    \hline
    \hline
    \noalign{\vskip 0.04in} 
    Disk size $(R_{90\%})$ $[\text{au}]$ & $M_{\star,\text{dyn}}$ $[M_{\odot}]$ & Uncertainty $[M_{\odot}]$ \\[0.4ex]
    \hline
    \hline
    \noalign{\vskip 0.04in} 
     & $M_\star\leq0.05$ & $0.03$ \\[0.4ex]
    $R_{90\%}\leq50$ & $0.05<M_\star\leq0.50$ & $0.15$ \\[0.4ex]
     & $M_\star>0.50$ & $0.30$ \\[0.4ex]
    \hline
    \noalign{\vskip 0.04in} 
     & $M_\star\leq0.05$ & $0.02$ \\[0.4ex]
    $50<R_{90\%}\leq90$ & $0.05<M_\star\leq0.50$ & $0.10$ \\[0.4ex]
     & $M_\star>0.50$ & $0.17$ \\[0.4ex]
    \hline
    \noalign{\vskip 0.04in} 
     & $M_\star\leq0.05$ & $0.03$ \\[0.4ex]
    $R_{90\%}>90$ & $0.05<M_\star\leq0.50$ & $0.15$ \\[0.4ex]
     & $M_\star>0.50$ & $0.11$ \\[0.4ex]
    \hline
    \hline
    \end{tabular}
    \caption{Table showing the logic used to assign the uncertainties to the dynamical stellar masses.}
    \label{table:M_*_uncertainty}
\end{table}

\section{HR diagrams}
\label{appendix:hrds}
In Fig. \ref{fig:hrd_spots_085}, we show the HR diagrams connected to the different evolutionary tracks used in this work. Since each model has a different validity based on different stellar ranges, some sources are found to fall outside (or very close to the edges of) the tracks. In these cases, it is not possible to derive masses and ages from HR diagram fitting using \verb|ysoisochrone|. We highlight in orange the sources for which it was impossible to derive mass and age, while we show in red the five sources we exclude from the analysis, as reported in Sect. \ref{sec:data}. The stellar masses are listed in Table \ref{Table:masses_total}.

\begin{figure*}[t]
    \centering
    \includegraphics[width=0.49\linewidth]{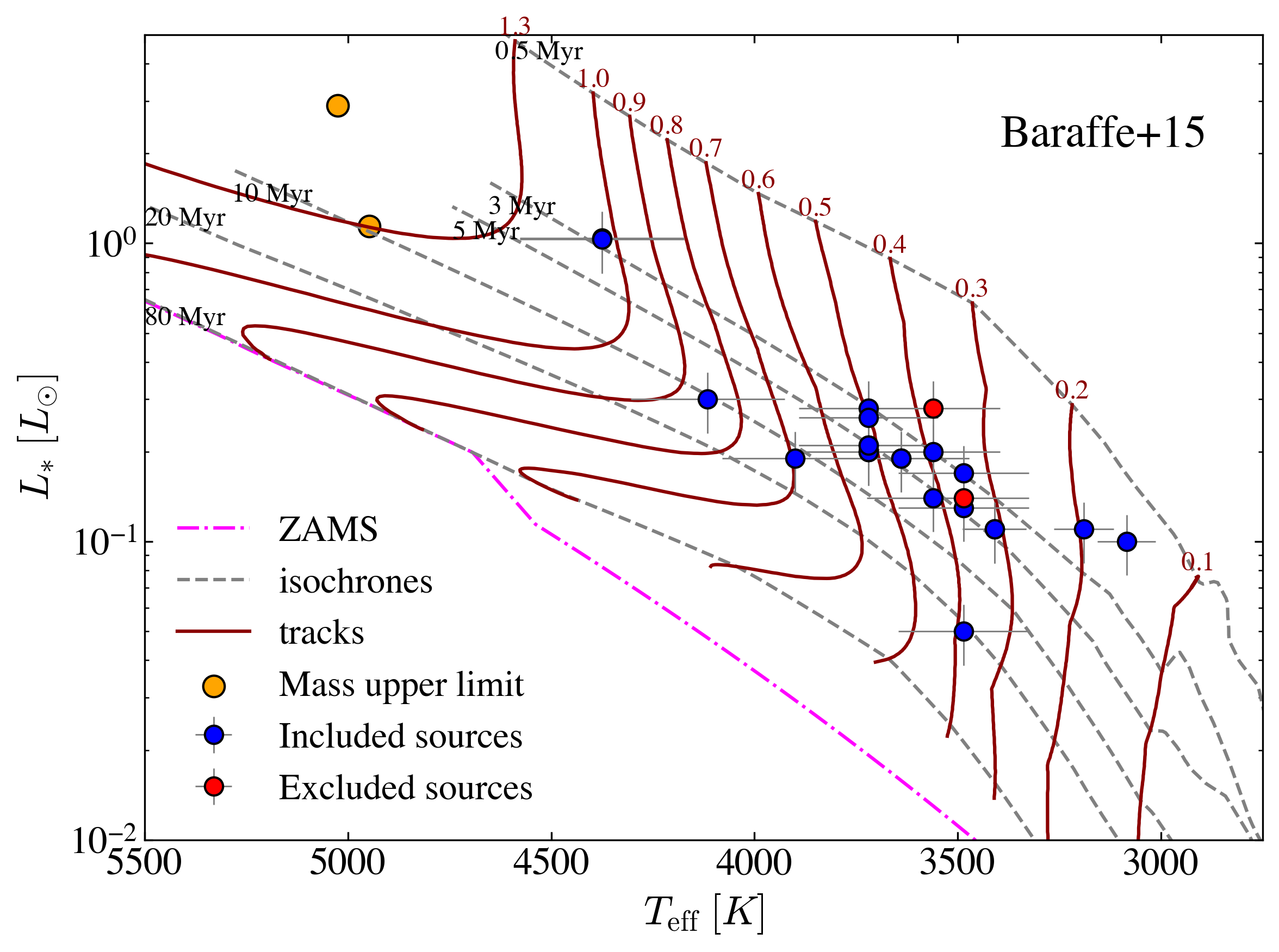}
    \includegraphics[width=0.49\linewidth]{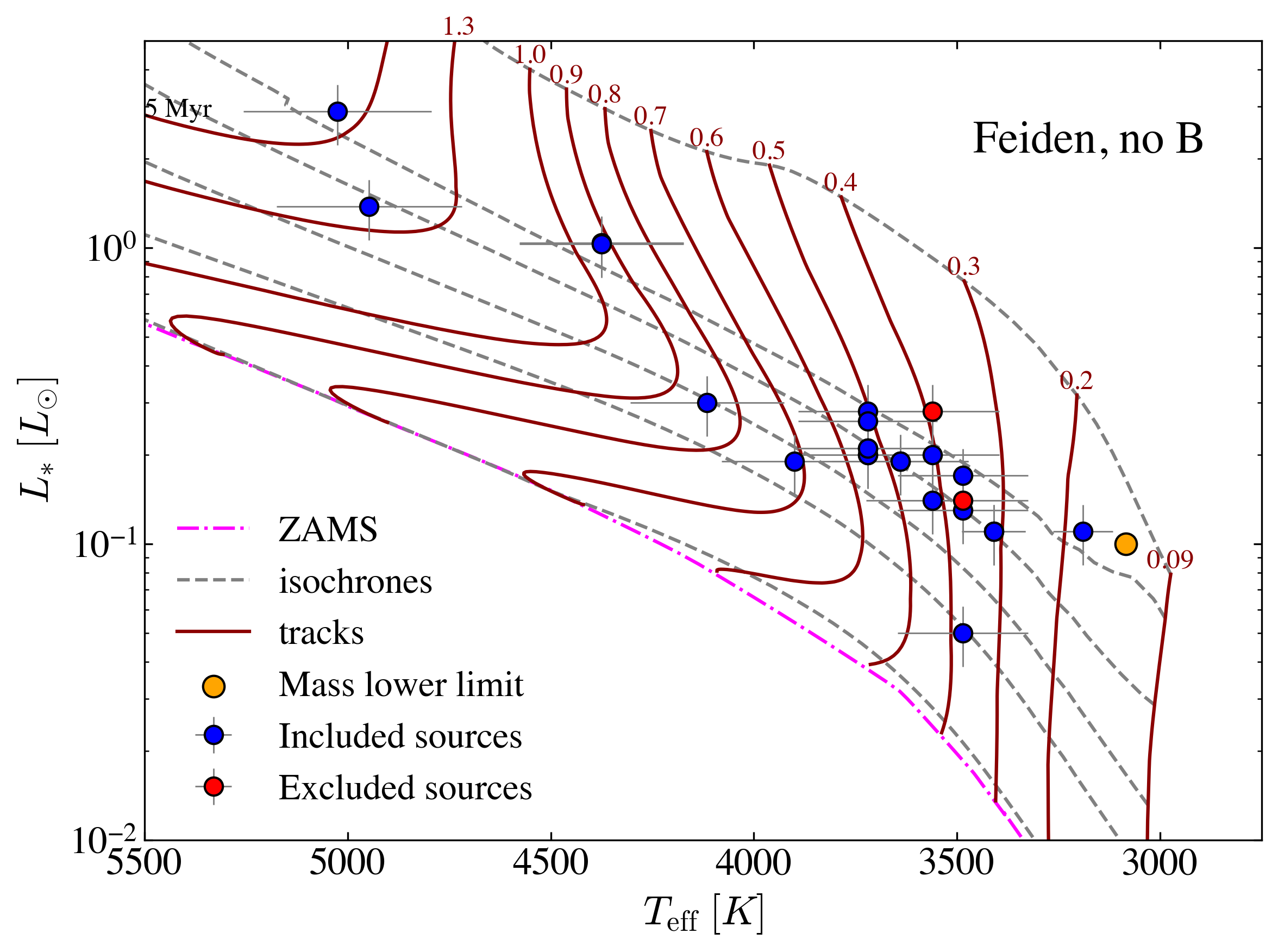}
    \includegraphics[width=0.49\linewidth]{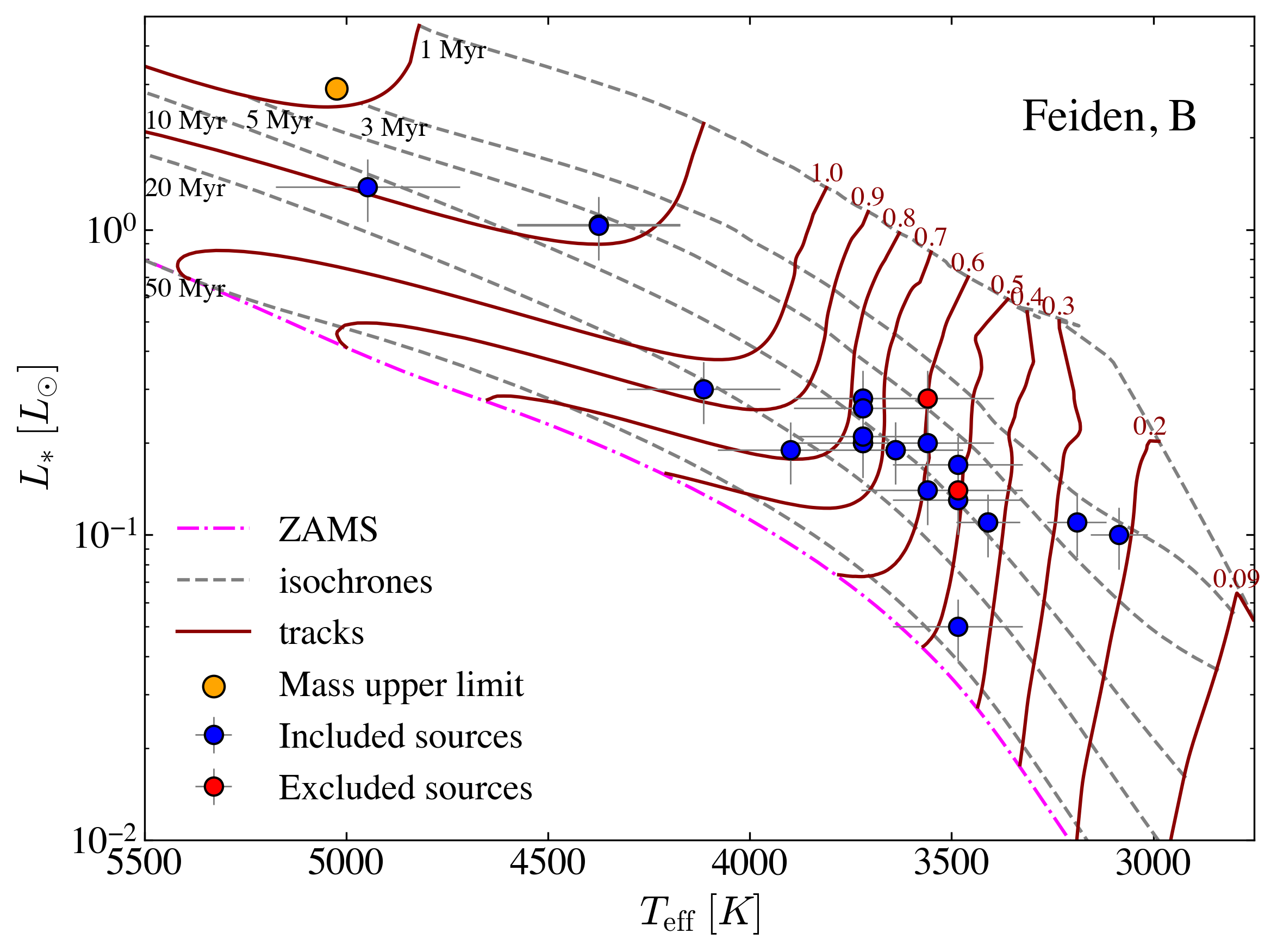}
    \includegraphics[width=0.49\linewidth]{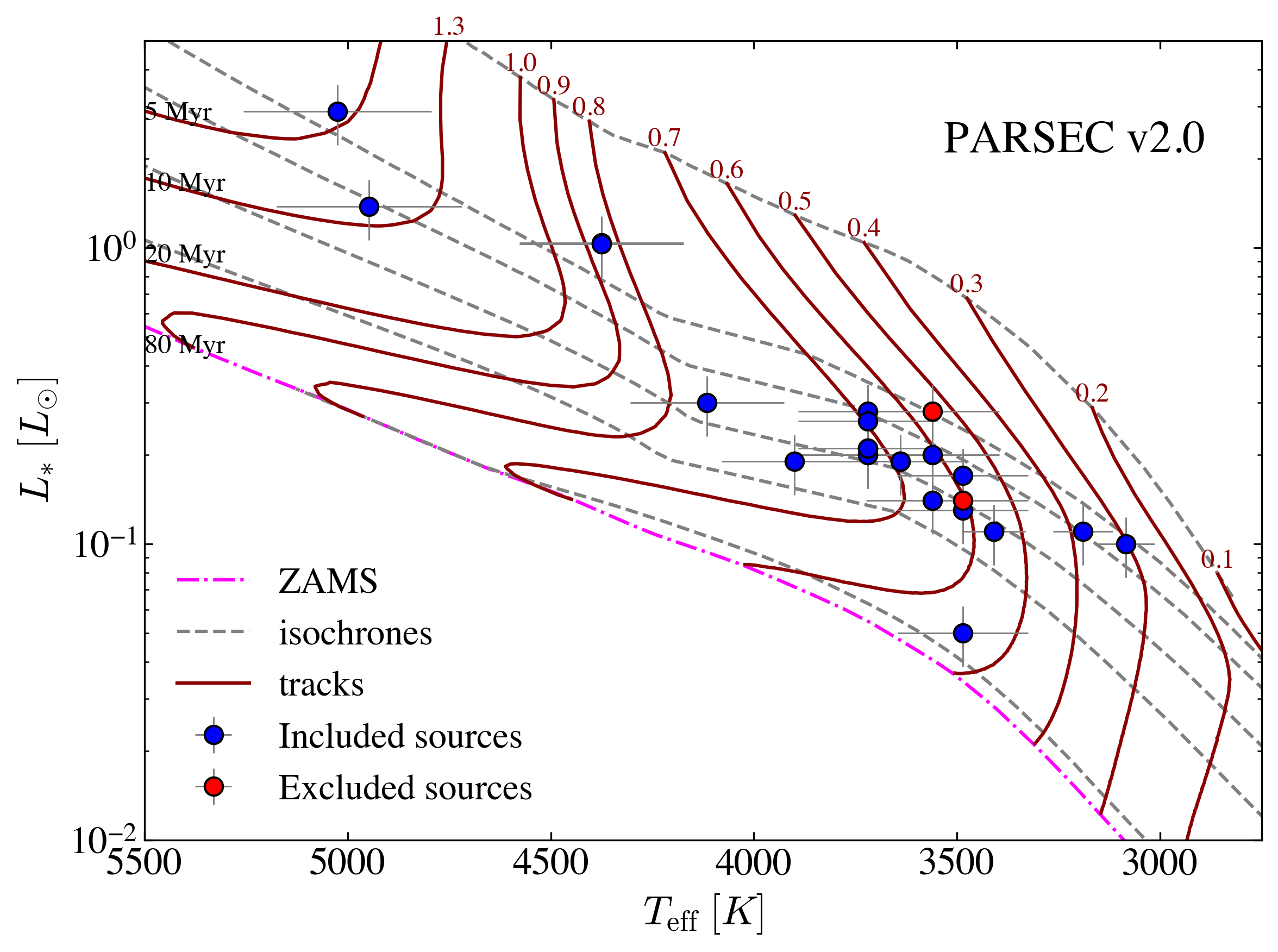}
    \includegraphics[width=0.49\linewidth]{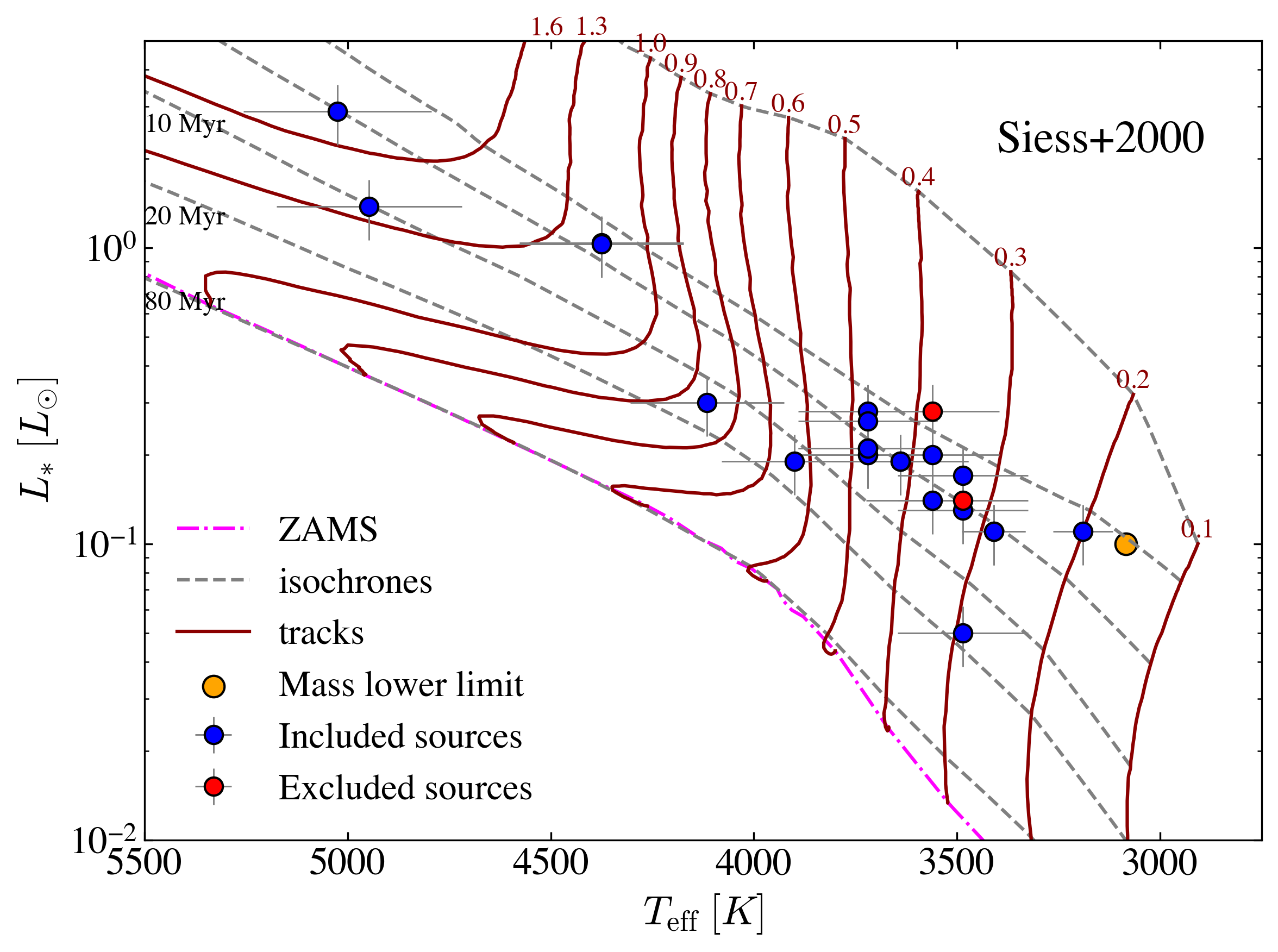}
    \includegraphics[width=0.49\linewidth]
    {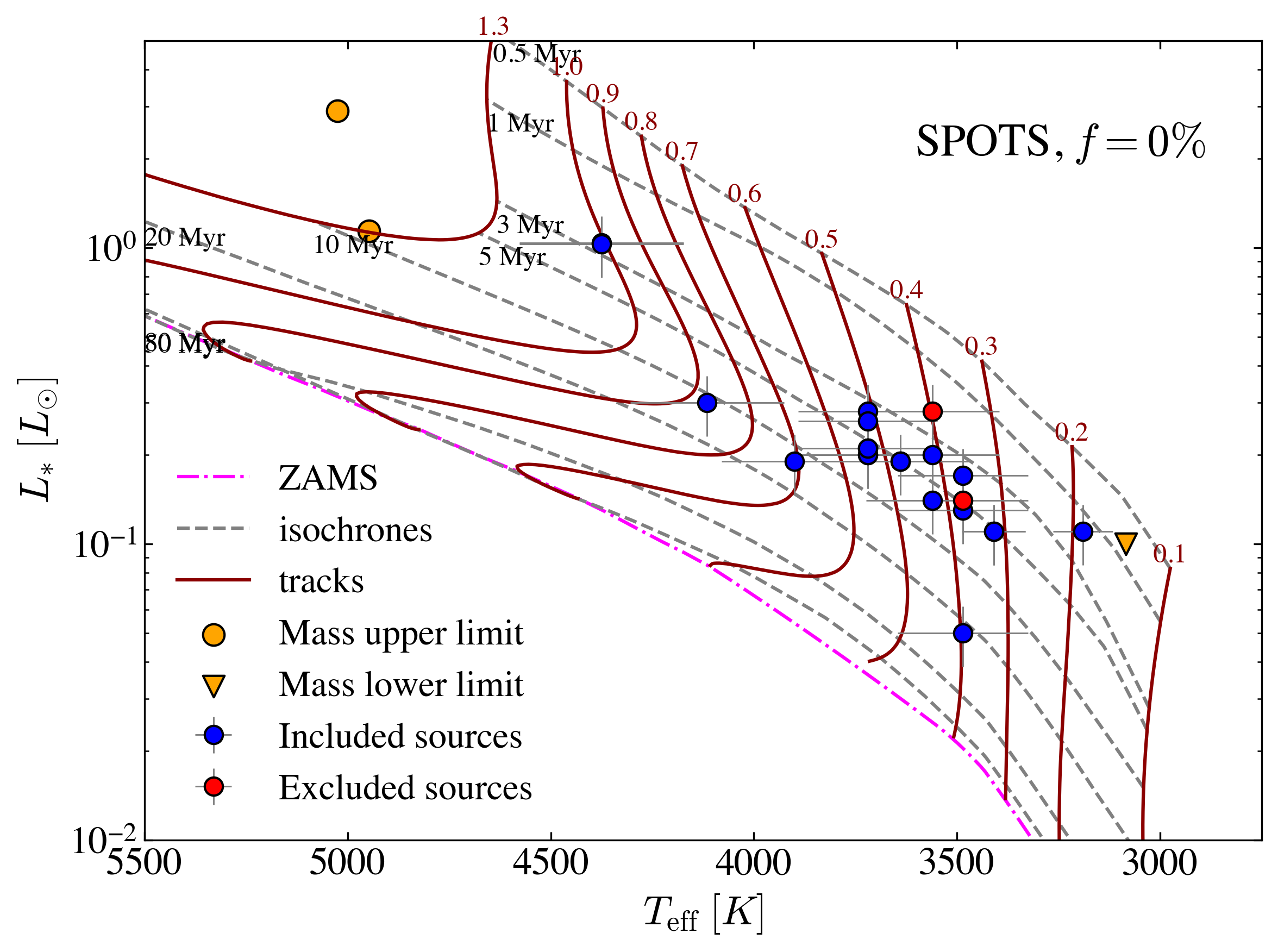}
    \caption{Hertzsprung–Russell diagrams with the different theoretical evolutionary tracks, the isochrones, and the 25 YSOs considered in this work.}
    \label{fig:hrd_spots_085}
\end{figure*}

\begin{figure*}[t]
    \ContinuedFloat
    \centering
    \includegraphics[width=0.49\linewidth]
    {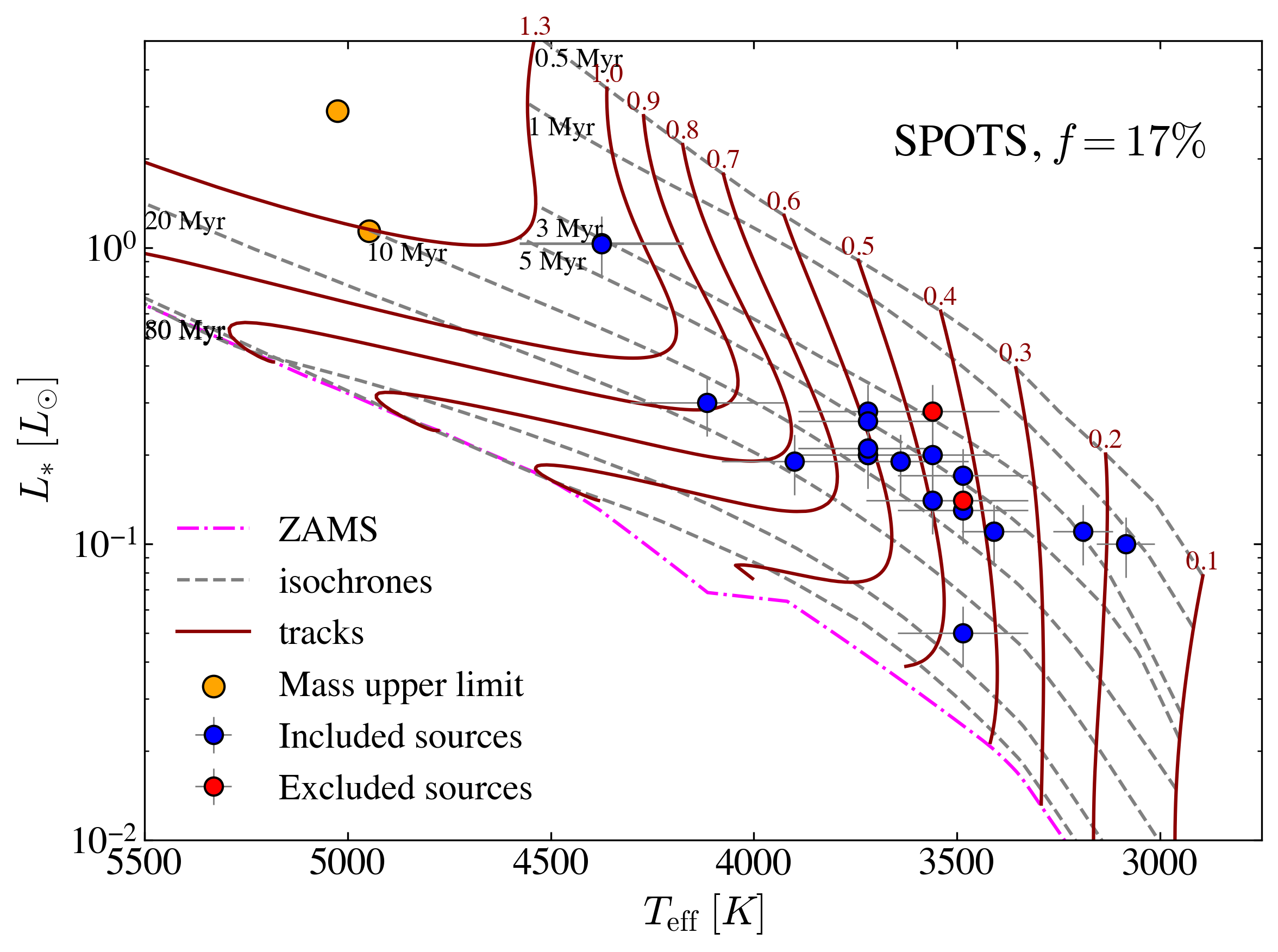}
    \includegraphics[width=0.49\linewidth]{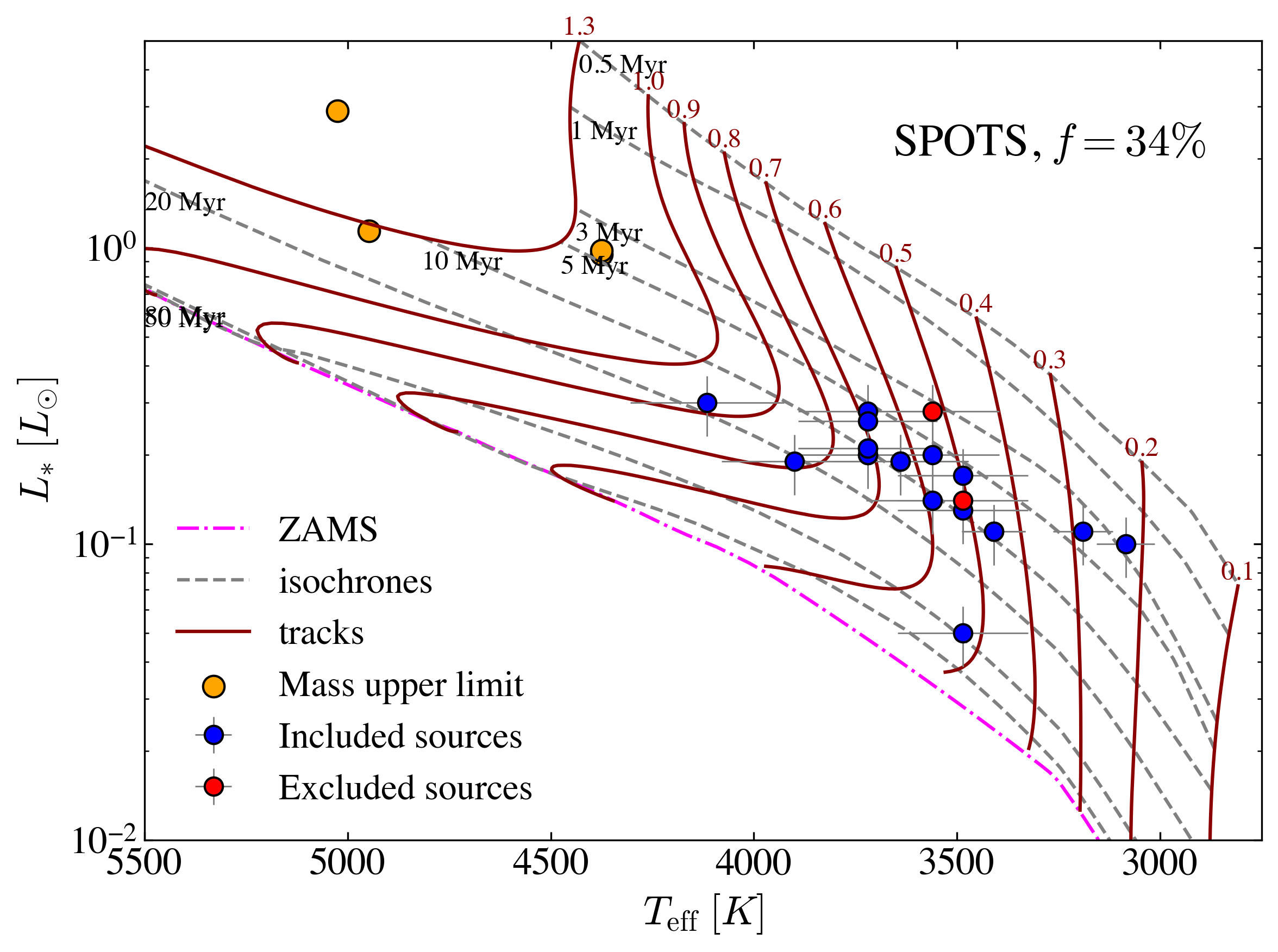}
    \includegraphics[width=0.49\linewidth]{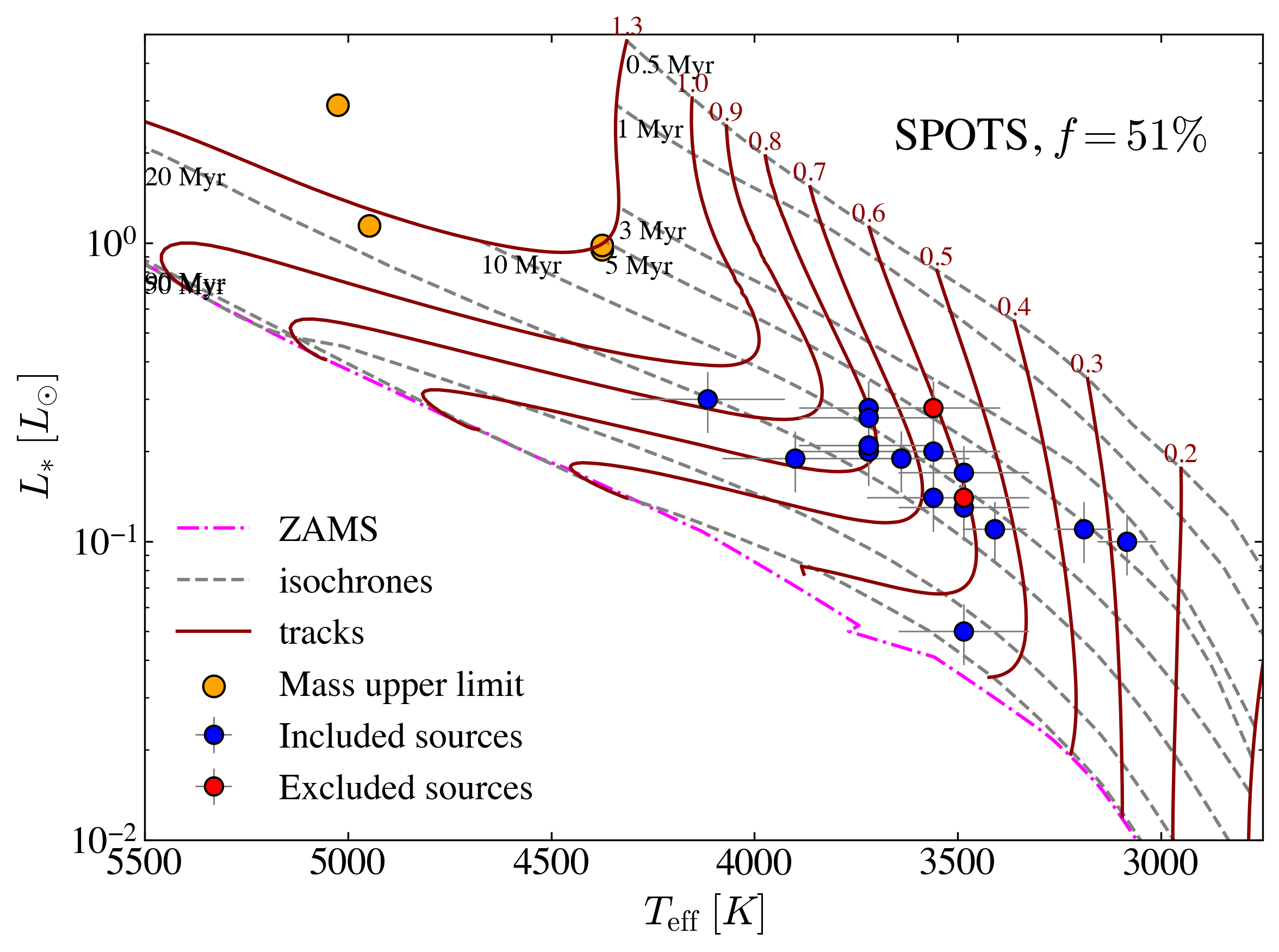}
    \includegraphics[width=0.49\linewidth]{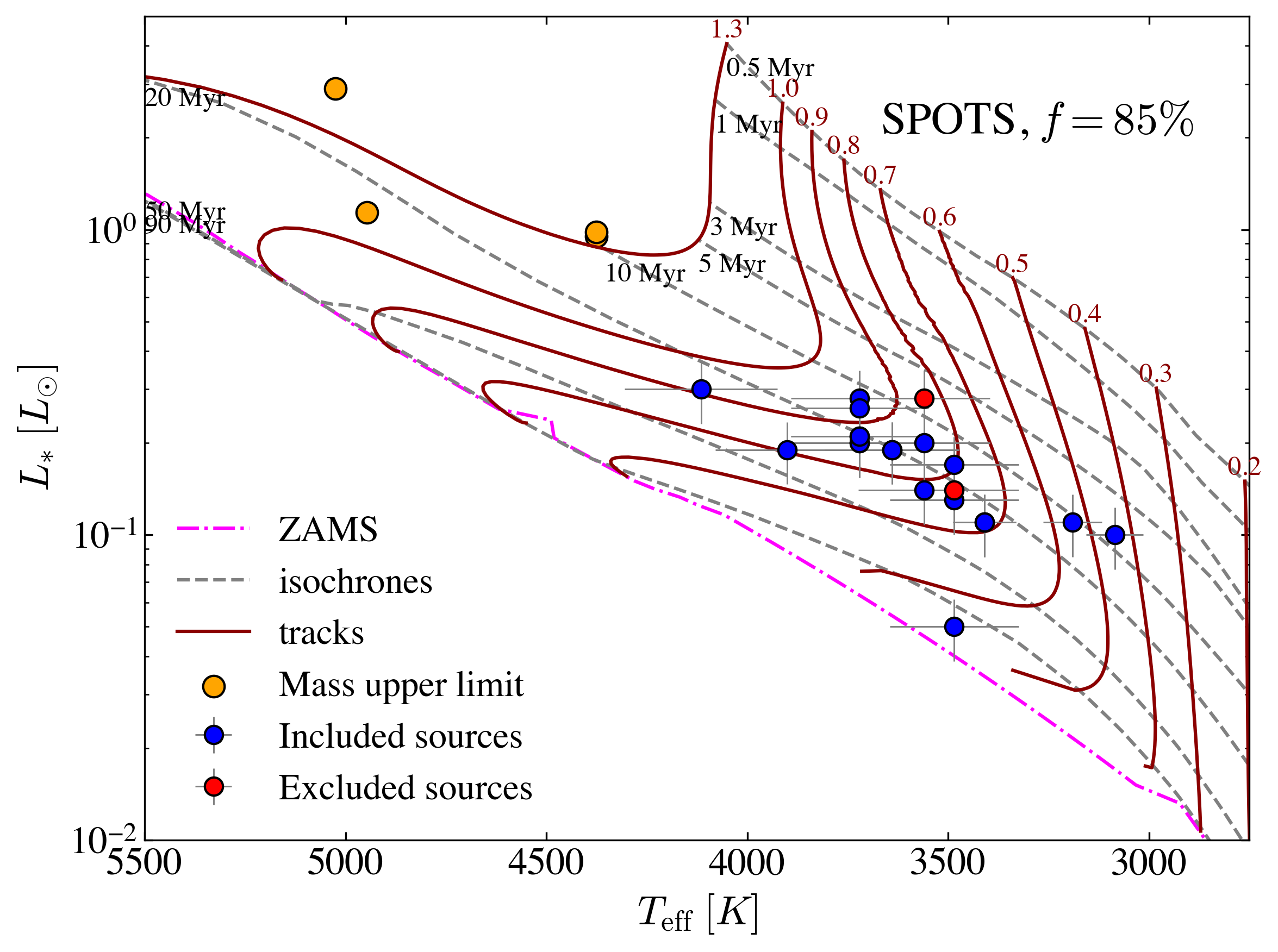}
    \caption{Continues from Fig. \ref{fig:hrd_spots_085} Hertzsprung–Russell diagrams with the different theoretical evolutionary tracks, the isochrones, and the 25 YSOs considered in this work.}
\end{figure*}

\section{Stellar ages and the consequences of a prior on the stellar mass}
\label{appendix:mass_priors}
\subsection{The impact of stellar mass priors on the age of single objects}
\begin{figure*}[t!!]
    \centering
    \includegraphics[width=0.49\linewidth]{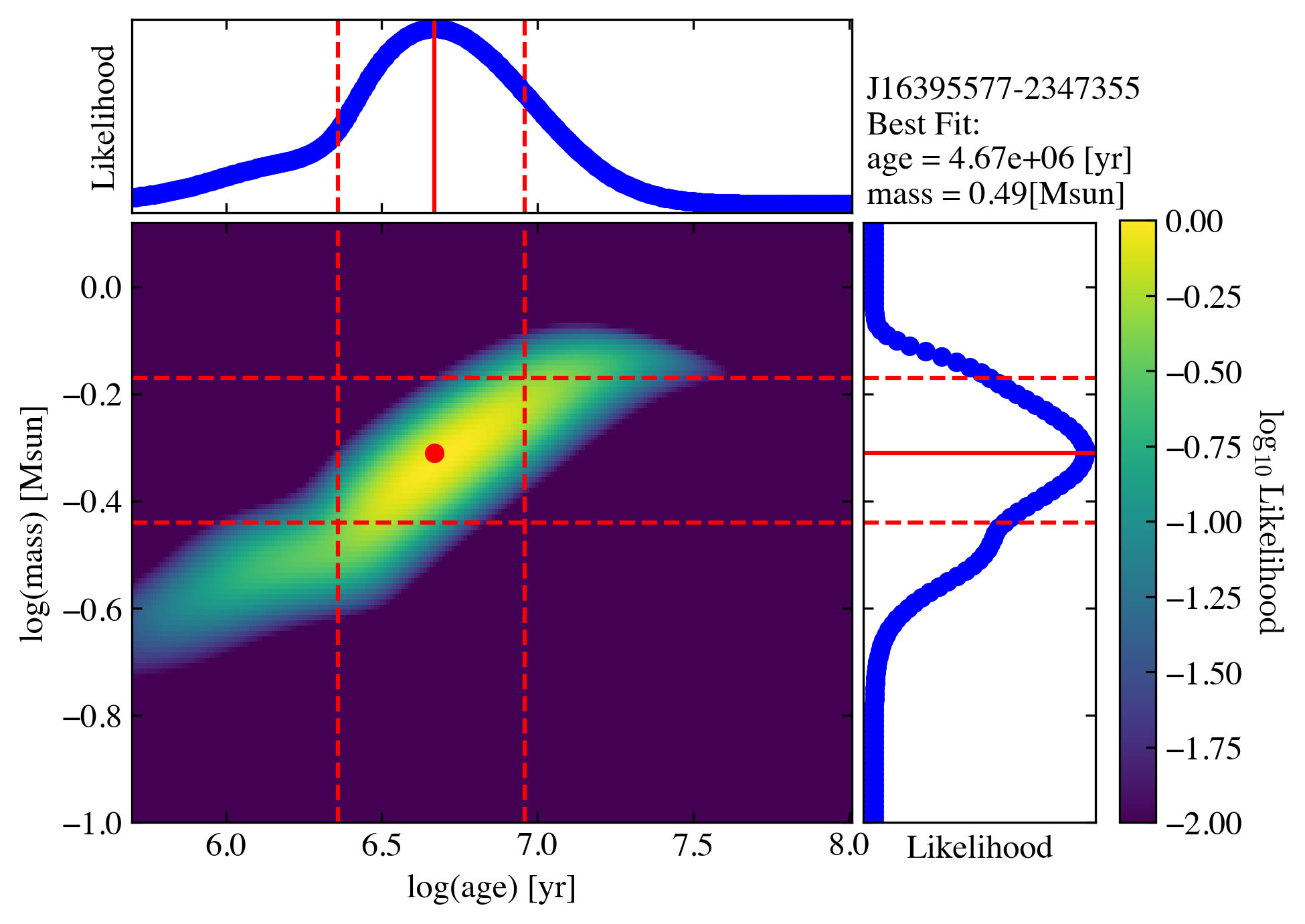}
    \includegraphics[width=0.49\linewidth]{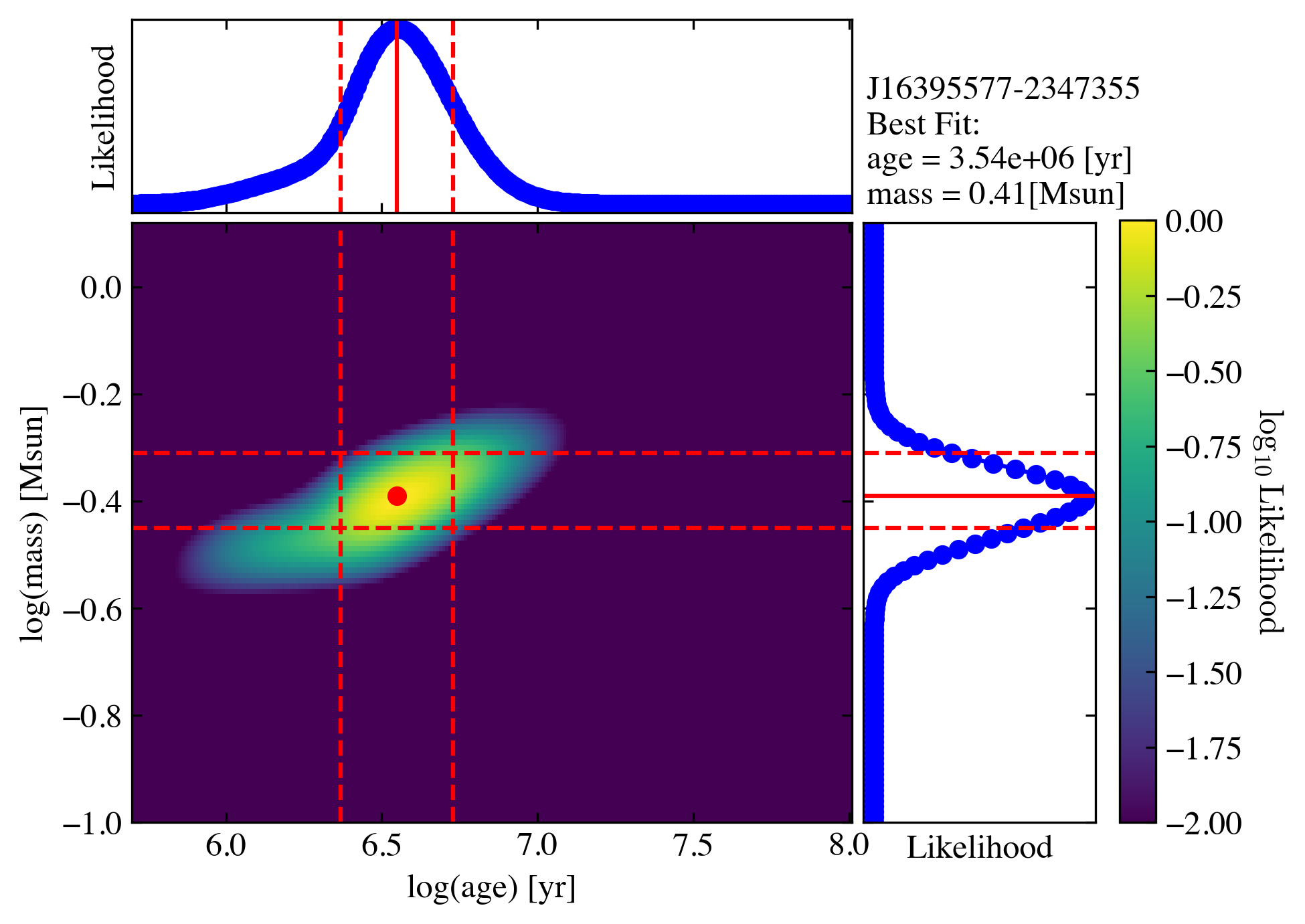}
    \caption{Example of age determination improvement by using dynamical stellar mass priors, using J16395577-2347355 as an example. On the left, we show the HR diagram best-fit posteriors of age (top panel) and mass (right panel). On the right, we show the distributions obtained after including a gaussian prior on the stellar mass built on the dynamical measure presented in \cite{Zallio_2026}. The red dot represents the peak of both the mass and age likelihoods.}
    \label{fig:mass_prior}
\end{figure*}

Using \verb|ysoisochrone|, we test how providing a prior on the stellar mass changes the results of HR diagram fitting (following what was done previously in the literature, e.g. \citealt{Rosenfeld_2012}). In Fig. \ref{fig:mass_prior} (left), we show the result of HR diagram fitting using the SPOTS $f=17\%$ tracks (\citealt{Somers_2020}) on one of the sources of our sample: the posterior distribution of age and mass is quite wide.
When including a gaussian mass prior based on the dynamical mass and its associated uncertainty, the age-mass degeneracy reduces greatly, as shown in Fig. \ref{fig:mass_prior} (right).
In the case of J16395577-2347355 (one of the most discrepant sources when considering SPOTS $f=17\%$) we show that the inferred median age changes by $\sim25\%$. When extending this analysis to the whole sample considered with SPOTS $f=17\%$, the median age changes from $6.3$ to $7$ Myr, as further shown in Figs. \ref{fig:ages}, \ref{fig:ages_prior}.

\subsection{Age distributions for different evolutionary tracks}
\label{appendix:ages}
\begin{figure*}[ht]
    \centering
    \includegraphics[width=0.75\textwidth]{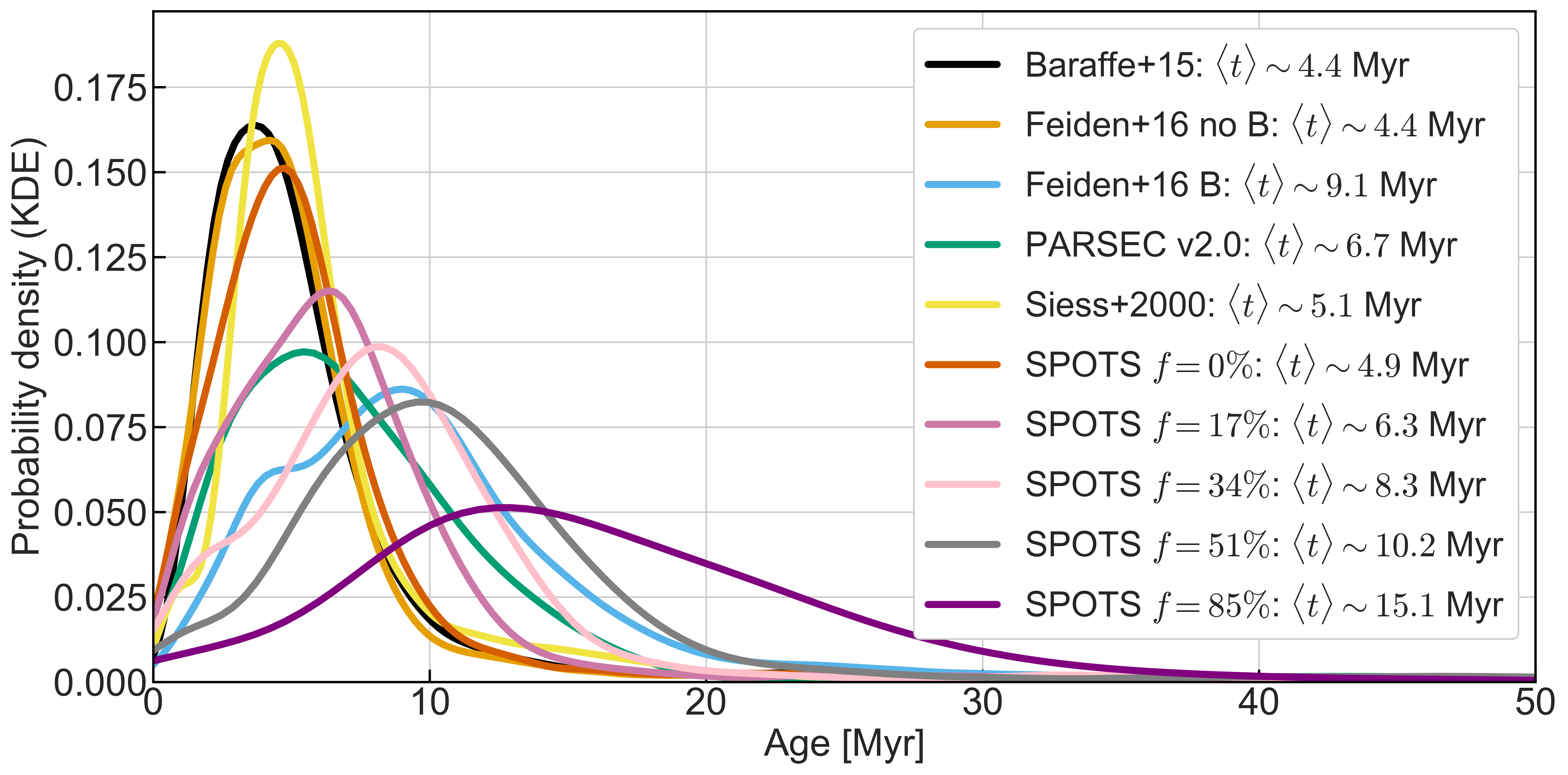}
    \caption{Different age distributions associated with different evolutionary tracks for the 20 Upper Scorpius sources considered for the comparison. The legend shows the median ages associated to each distribution.}
    \label{fig:ages}
\end{figure*}

In Fig. \ref{fig:ages}, we present the age distributions derived from the HR diagram fits performed with \verb|ysoisochrone|, without applying any mass priors. These distributions are based on the posterior ages obtained for each source. The visualization was produced using the \verb|seaborn| package (\citealt{Waskom2021}), which implements Kernel Density Estimation (KDE)\footnote{The Kernel Density Estimation (KDE) is a non-parametric method used to estimate the probability density function (PDF) of a continuous variable from a finite sample.} to provide a smooth representation of the underlying probability density. The resulting distributions show substantial differences among the ten evolutionary models considered. The median age across the models is $\sim6.5$ Myr, with a scatter of $\sim3.4$ Myr.

\begin{figure*}[ht]
    \centering
    \includegraphics[width=0.75\textwidth]{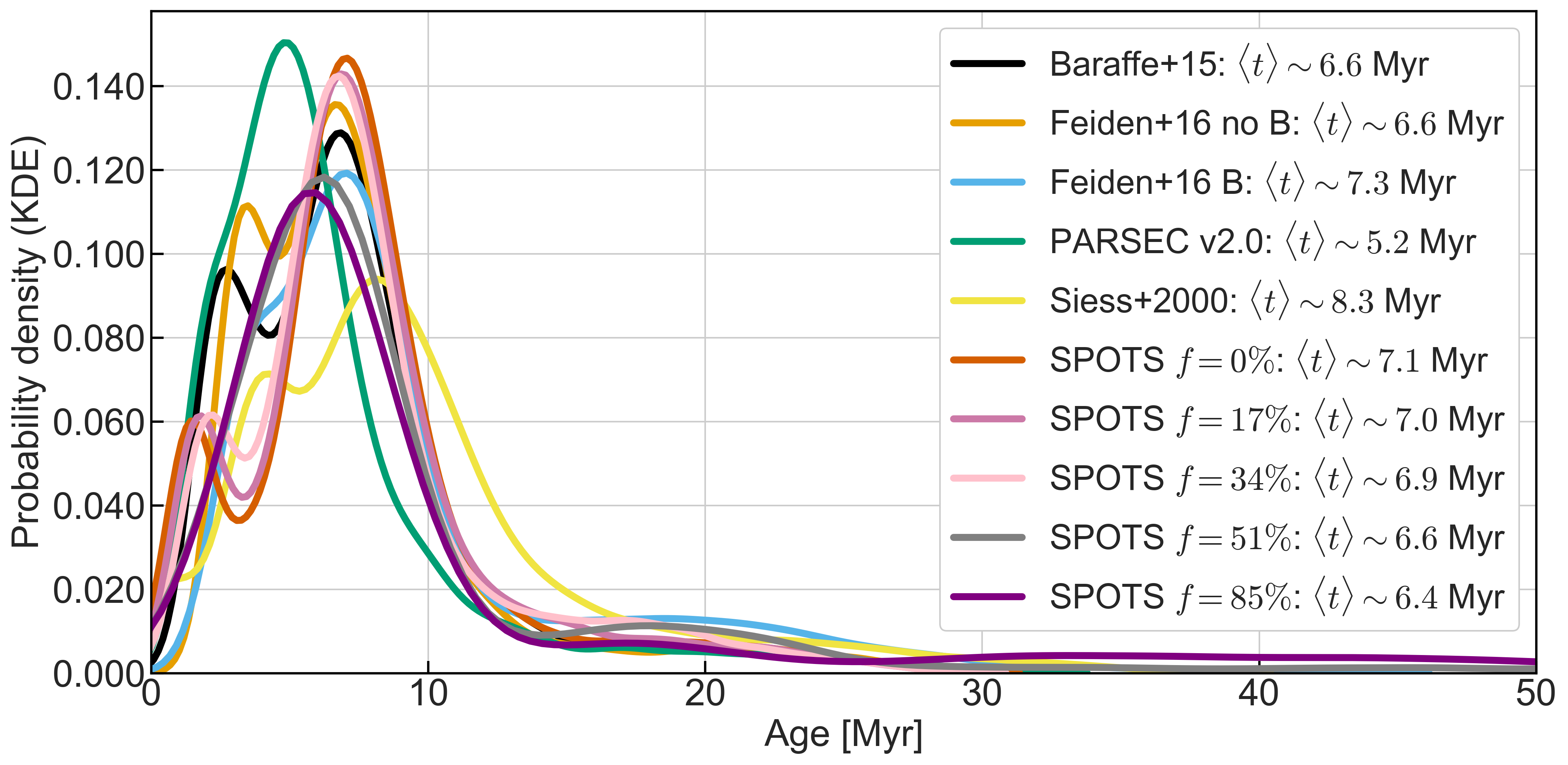}
    \caption{Different age distributions associated with different evolutionary tracks for the 20 Upper Scorpius sources considered for the comparison after fitting the single HR diagrams using mass priors coming from the stellar dynamical masses. The legend shows the median ages associated to each distribution.}
    \label{fig:ages_prior}
\end{figure*}

Figure \ref{fig:ages_prior} displays the corresponding age distributions obtained after including priors on the stellar dynamical masses. When the mass prior is introduced, the differences among the distributions become notably smaller. The median age across the ten evolutionary tracks increases slightly to $\sim6.8$ Myr, while the scatter between the tracks decreases to $\sim0.8$ Myr.

This result can be explained as follows. The dominant timescale for pre-main sequence stars is the Kelvin-Helmholtz timescale\footnote{The Kelvin-Helmholtz timescale is the time over which a star radiates away its gravitational potential energy at its current luminosity: $\tau_{KH} \sim GM_\star^2 / R_\star L_\star$, where $G$ is the gravitational constant, $M_\star$ the mass of the star, $R_\star$ its radius and $L_\star$ its luminosity.}, since the nuclear fusion has not started yet (see e.g., \citealt{Hartmann_1998}). The stellar radius is fixed by knowing both $T_{eff}$ and $L_\star$, thus when we consider a prior on the stellar masses, we force the ages to converge to a fixed value.

\end{appendix}

\begin{landscape}
\begin{table}
\centering
\resizebox{\linewidth}{!}{
\setcounter{table}{0}
\centering
\begin{tabular}{l c c c c c c c c c c c}
\hline
\hline
\noalign{\vskip 0.03in} 
Source (2MASS) & $M_{\star,\texttt{dyn}}$ [$M_{\odot}$] & Baraffe+15 [$M_{\odot}$] & Feiden+16, no B [$M_{\odot}$] & Feiden+16, B [$M_{\odot}$] & PARSEC v2.0 [$M_{\odot}$] & Siess+2000 [$M_{\odot}$] & SPOTS $f=0\%$ [$M_{\odot}$] & SPOTS $f=17\%$ [$M_{\odot}$] & SPOTS $f=34\%$ [$M_{\odot}$] & SPOTS $f=51\%$ [$M_{\odot}$] & SPOTS $f=85\%$ [$M_{\odot}$] \\ [0.2ex]
\hline
\hline
\noalign{\vskip 0.03in}
J15583692-2257153 & $0.67 \pm 0.11$ & $>1.30$ & $1.64^{+0.47}_{-0.18}$ & $>1.70$ & $1.60^{+0.46}_{-0.17}$ & $1.74^{+0.21}_{-0.15}$ & $>1.30$ & $>1.30$ & $>1.30$ & $>1.30$ & $>1.30$ \\ [0.4ex]
J16035793-1942108 & $0.52 \pm 0.10$ & $0.42^{+0.15}_{-0.11}$ & $0.43^{+0.15}_{-0.10}$ & $0.64^{+0.19}_{-0.12}$ & $0.67^{+0.06}_{-0.09}$ & $0.37^{+0.11}_{-0.08}$ & $0.44^{+0.17}_{-0.12}$ & $0.51^{+0.16}_{-0.11}$ & $0.66^{+0.15}_{-0.12}$ & $0.69^{+0.16}_{-0.10}$ & $0.76^{+0.15}_{-0.07}$ \\ [0.4ex]
J16052157-1821412 & $1.12 \pm 0.11$ & $1.12^{+0.26}_{-0.19}$ & $0.96^{+0.22}_{-0.18}$ & $1.32^{+0.20}_{-0.09}$ & $0.88^{+0.18}_{-0.15}$ & $1.20^{+0.31}_{-0.16}$ & $1.12^{+0.17}_{-0.19}$ & $1.17^{+0.11}_{-0.15}$ & $>1.30$ & $>1.30$ & $>1.30$ \\ [0.4ex]
J16062861-2121297 & $0.53 \pm 0.10$ & $0.47^{+0.16}_{-0.11}$ & $0.47^{+0.15}_{-0.11}$ & $0.72^{+0.21}_{-0.12}$ & $0.70^{+0.05}_{-0.08}$ & $0.43^{+0.12}_{-0.10}$ & $0.49^{+0.19}_{-0.13}$ & $0.56^{+0.16}_{-0.13}$ & $0.72^{+0.17}_{-0.12}$ & $0.76^{+0.17}_{-0.10}$ & $0.81^{+0.10}_{-0.07}$ \\ [0.4ex]
J16095933-1800090 & $0.18 \pm 0.15$ & $0.20^{+0.04}_{-0.03}$ & $0.19^{+0.10}_{-0.04}$ & $0.28^{+0.05}_{-0.04}$ & $0.38^{+0.12}_{-0.06}$ & $0.20^{+0.04}_{-0.03}$ & $0.19^{+0.04}_{-0.03}$ & $0.22^{+0.04}_{-0.03}$ & $0.26^{+0.05}_{-0.04}$ & $0.35^{+0.08}_{-0.05}$ & $0.55^{+0.08}_{-0.07}$ \\ [0.4ex]
J16101264-2104446 & $1.23 \pm 0.11$ & $1.12^{+0.26}_{-0.19}$ & $0.96^{+0.22}_{-0.18}$ & $1.32^{+0.20}_{-0.09}$ & $0.86^{+0.17}_{-0.14}$ & $1.20^{+0.31}_{-0.16}$ & $1.12^{+0.17}_{-0.19}$ & $1.17^{+0.11}_{-0.15}$ & $>1.30$ & $>1.30$ & $>1.30$ \\ [0.4ex]
J16123916-1859284 & $0.60 \pm 0.11$ & $0.52^{+0.17}_{-0.12}$ & $0.54^{+0.16}_{-0.12}$ & $0.76^{+0.20}_{-0.10}$ & $0.71^{+0.09}_{-0.06}$ & $0.48^{+0.12}_{-0.11}$ & $0.55^{+0.16}_{-0.12}$ & $0.71^{+0.18}_{-0.13}$ & $0.76^{+0.15}_{-0.11}$ & $0.78^{+0.18}_{-0.08}$ & $0.83^{+0.08}_{-0.06}$ \\ [0.4ex]
J16140792-1938292 & $1.20 \pm 0.17$ & $>1.30$ & $1.33^{+0.16}_{-0.12}$ & $1.48^{+0.22}_{-0.13}$ & $1.30^{+0.16}_{-0.11}$ & $1.38^{+0.13}_{-0.12}$ & $>1.30$ & $>1.30$ & $>1.30$ & $>1.30$ & $>1.30$ \\ [0.4ex]
J16145024-2100599 & $0.67 \pm 0.11$ & $0.50^{+0.16}_{-0.11}$ & $0.51^{+0.15}_{-0.11}$ & $0.81^{+0.21}_{-0.12}$ & $0.70^{+0.14}_{-0.09}$ & $0.49^{+0.14}_{-0.10}$ & $0.52^{+0.18}_{-0.13}$ & $0.60^{+0.19}_{-0.13}$ & $0.78^{+0.20}_{-0.15}$ & $0.83^{+0.19}_{-0.12}$ & $0.89^{+0.11}_{-0.08}$ \\ [0.4ex]
J16152752-1847097 & $0.59 \pm 0.11$ & $0.52^{+0.17}_{-0.12}$ & $0.54^{+0.16}_{-0.12}$ & $0.78^{+0.20}_{-0.10}$ & $0.71^{+0.09}_{-0.06}$ & $0.48^{+0.12}_{-0.11}$ & $0.55^{+0.17}_{-0.12}$ & $0.71^{+0.18}_{-0.13}$ & $0.76^{+0.17}_{-0.11}$ & $0.79^{+0.16}_{-0.09}$ & $0.85^{+0.17}_{-0.06}$ \\ [0.4ex]
J16181445-2319251 & $0.21 \pm 0.15$ & $0.14^{+0.02}_{-0.02}$ & $<0.10$ & $0.22^{+0.04}_{-0.03}$ & $0.32^{+0.15}_{-0.07}$ & $<0.10$ & $<0.10$ & $0.19^{+0.04}_{-0.03}$ & $0.22^{+0.04}_{-0.03}$ & $0.26^{+0.04}_{-0.04}$ & $0.45^{+0.09}_{-0.07}$ \\ [0.4ex]
J16202291-2227041 & $0.39 \pm 0.10$ & $0.32^{+0.06}_{-0.05}$ & $0.32^{+0.06}_{-0.05}$ & $0.44^{+0.08}_{-0.06}$ & $0.55^{+0.08}_{-0.06}$ & $0.30^{+0.04}_{-0.03}$ & $0.32^{+0.07}_{-0.05}$ & $0.39^{+0.08}_{-0.06}$ & $0.46^{+0.08}_{-0.06}$ & $0.56^{+0.10}_{-0.06}$ & $0.71^{+0.10}_{-0.03}$ \\ [0.4ex]
J16202863-2442087 & $0.72 \pm 0.11$ & $0.51^{+0.16}_{-0.11}$ & $0.52^{+0.15}_{-0.12}$ & $0.79^{+0.23}_{-0.12}$ & $0.70^{+0.14}_{-0.09}$ & $0.49^{+0.14}_{-0.10}$ & $0.52^{+0.18}_{-0.14}$ & $0.60^{+0.17}_{-0.13}$ & $0.78^{+0.20}_{-0.13}$ & $0.83^{+0.17}_{-0.11}$ & $0.89^{+0.18}_{-0.06}$ \\ [0.4ex]
J16203960-2634284 & $0.55 \pm 0.11$ & $0.47^{+0.16}_{-0.11}$ & $0.47^{+0.15}_{-0.11}$ & $0.72^{+0.21}_{-0.12}$ & $0.70^{+0.05}_{-0.08}$ & $0.43^{+0.12}_{-0.10}$ & $0.49^{+0.19}_{-0.13}$ & $0.56^{+0.16}_{-0.13}$ & $0.72^{+0.17}_{-0.12}$ & $0.76^{+0.17}_{-0.10}$ & $0.81^{+0.10}_{-0.07}$ \\ [0.4ex]
J16215472-2752053 & $0.84 \pm 0.11$ & $0.83^{+0.19}_{-0.11}$ & $0.84^{+0.17}_{-0.09}$ & $0.91^{+0.09}_{-0.08}$ & $0.77^{+0.05}_{-0.07}$ & $0.83^{+0.22}_{-0.11}$ & $0.85^{+0.17}_{-0.09}$ & $0.87^{+0.18}_{-0.08}$ & $0.89^{+0.11}_{-0.08}$ & $0.91^{+0.09}_{-0.06}$ & $0.95^{+0.22}_{-0.08}$ \\ [0.4ex]
J16221532-2511349 & $0.47 \pm 0.15$ & $0.37^{+0.14}_{-0.10}$ & $0.38^{+0.14}_{-0.10}$ & $0.52^{+0.15}_{-0.12}$ & $0.64^{+0.08}_{-0.09}$ & $0.34^{+0.10}_{-0.07}$ & $0.39^{+0.21}_{-0.11}$ & $0.45^{+0.17}_{-0.12}$ & $0.54^{+0.17}_{-0.12}$ & $0.66^{+0.17}_{-0.11}$ & $0.72^{+0.11}_{-0.06}$ \\ [0.4ex]
J16230761-2516339 & $0.37 \pm 0.10$ & $0.39^{+0.12}_{-0.10}$ & $0.40^{+0.19}_{-0.10}$ & $0.47^{+0.16}_{-0.09}$ & $0.53^{+0.12}_{-0.05}$ & $0.30^{+0.10}_{-0.08}$ & $0.43^{+0.16}_{-0.10}$ & $0.49^{+0.13}_{-0.09}$ & $0.52^{+0.11}_{-0.07}$ & $0.55^{+0.11}_{-0.05}$ & $0.58^{+0.04}_{-0.04}$ \\ [0.4ex]
J16253798-1943162 & $0.72 \pm 0.11$ & $0.72^{+0.17}_{-0.11}$ & $0.73^{+0.15}_{-0.09}$ & $0.78^{+0.09}_{-0.07}$ & $0.71^{+0.11}_{-0.05}$ & $0.69^{+0.20}_{-0.12}$ & $0.72^{+0.17}_{-0.09}$ & $0.76^{+0.15}_{-0.08}$ & $0.78^{+0.18}_{-0.07}$ & $0.79^{+0.08}_{-0.07}$ & $0.83^{+0.10}_{-0.07}$ \\ [0.4ex]
J16293267-2543291 & $0.45 \pm 0.10$ & $0.35^{+0.12}_{-0.09}$ & $0.38^{+0.14}_{-0.10}$ & $0.52^{+0.15}_{-0.12}$ & $0.57^{+0.16}_{-0.12}$ & $0.35^{+0.09}_{-0.07}$ & $0.28^{+0.10}_{-0.13}$ & $0.44^{+0.19}_{-0.12}$ & $0.51^{+0.18}_{-0.12}$ & $0.68^{+0.19}_{-0.14}$ & $0.78^{+0.16}_{-0.08}$ \\ [0.4ex]
J16395577-2347355 & $0.38 \pm 0.15$ & $0.40^{+0.14}_{-0.10}$ & $0.41^{+0.13}_{-0.10}$ & $0.62^{+0.18}_{-0.13}$ & $0.62^{+0.11}_{-0.10}$ & $0.38^{+0.10}_{-0.09}$ & $0.43^{+0.22}_{-0.11}$ & $0.49^{+0.19}_{-0.13}$ & $0.56^{+0.16}_{-0.14}$ & $0.74^{+0.17}_{-0.12}$ & $0.81^{+0.16}_{-0.09}$ \\ [0.4ex]

\hline
\hline
\noalign{\vskip 0.03in} 
Excluded (red; 2MASS) & $M_{\star,\texttt{dyn}}$ [$M_{\odot}$] & Baraffe+15 [$M_{\odot}$] & Feiden+16, no B [$M_{\odot}$] & Feiden+16, B [$M_{\odot}$] & PARSEC v2.0 [$M_{\odot}$] & Siess+2000 [$M_{\odot}$] & SPOTS $f=0\%$ [$M_{\odot}$] & SPOTS $f=17\%$ [$M_{\odot}$] & SPOTS $f=34\%$ [$M_{\odot}$] & SPOTS $f=51\%$ [$M_{\odot}$] & SPOTS $f=85\%$ [$M_{\odot}$] \\[0.2ex]
\hline
\hline
\noalign{\vskip 0.03in}
J15583620-1946135 & $0.07 \pm 0.15$ & $0.32^{+0.06}_{-0.05}$ & $0.32^{+0.06}_{-0.05}$ & $0.44^{+0.08}_{-0.06}$ & $0.55^{+0.08}_{-0.06}$ & $0.30^{+0.04}_{-0.03}$ & $0.32^{+0.07}_{-0.05}$ & $0.39^{+0.08}_{-0.06}$ & $0.46^{+0.08}_{-0.06}$ & $0.56^{+0.10}_{-0.06}$ & $0.71^{+0.10}_{-0.03}$ \\ [0.4ex]
J16132190-2136136 & $0.59 \pm 0.30$ & $0.36^{+0.14}_{-0.09}$ & $0.38^{+0.14}_{-0.10}$ & $0.51^{+0.15}_{-0.11}$ & $0.64^{+0.08}_{-0.11}$ & $0.34^{+0.10}_{-0.07}$ & $0.38^{+0.20}_{-0.11}$ & $0.45^{+0.18}_{-0.12}$ & $0.52^{+0.17}_{-0.13}$ & $0.68^{+0.16}_{-0.11}$ & $0.74^{+0.15}_{-0.07}$ \\ [0.4ex]
J16222982-2002472 & $1.37 \pm 0.17$ & $0.38^{+0.12}_{-0.09}$ & $0.41^{+0.14}_{-0.10}$ & $0.60^{+0.17}_{-0.13}$ & $0.55^{+0.18}_{-0.11}$ & $0.40^{+0.10}_{-0.08}$ & $0.35^{+0.10}_{-0.08}$ & $0.37^{+0.12}_{-0.15}$ & $0.54^{+0.20}_{-0.13}$ & $0.62^{+0.18}_{-0.15}$ & $0.87^{+0.18}_{-0.11}$ \\ [0.4ex]
J16271273-2504017 & $1.15 \pm 0.17$ & $0.52^{+0.17}_{-0.12}$ & $0.54^{+0.16}_{-0.12}$ & $0.76^{+0.20}_{-0.10}$ & $0.71^{+0.09}_{-0.06}$ & $0.48^{+0.12}_{-0.11}$ & $0.55^{+0.16}_{-0.12}$ & $0.71^{+0.18}_{-0.13}$ & $0.76^{+0.15}_{-0.11}$ & $0.78^{+0.18}_{-0.08}$ & $0.83^{+0.08}_{-0.06}$ \\ [0.4ex]
J16274905-2602437 & $0.20 \pm 0.10$ & $0.52^{+0.17}_{-0.12}$ & $0.54^{+0.16}_{-0.12}$ & $0.78^{+0.20}_{-0.10}$ & $0.71^{+0.09}_{-0.06}$ & $0.48^{+0.12}_{-0.11}$ & $0.55^{+0.17}_{-0.12}$ & $0.71^{+0.18}_{-0.13}$ & $0.76^{+0.17}_{-0.11}$ & $0.79^{+0.16}_{-0.09}$ & $0.85^{+0.17}_{-0.06}$ \\ [0.4ex]
\hline
\hline
\end{tabular}
}
\vspace{3pt}
\caption{Table containing the stellar masses considered in this work. The disk-based dynamical masses were taken from \cite{Zallio_2026}. The full table is available in the online material in machine-readable form and on \texttt{GitHub}\hyperref[footnote:github]{\textsuperscript{\ref{footnote:github}}}.}
\vspace{-0.3cm}
\label{Table:masses_total}
\end{table}

\end{landscape}
\clearpage

\end{document}